%% file: acl_latex.tex
\pdfoutput=1

\documentclass[11pt]{article}

\usepackage[final]{acl}

\usepackage{times}
\usepackage{latexsym}

\usepackage[T1]{fontenc}

\usepackage[utf8]{inputenc}

\usepackage{microtype}

\usepackage{inconsolata}

\usepackage{graphicx}
\usepackage{booktabs}
\usepackage{algorithm}
\usepackage{algorithmic}
\usepackage{hyperref}
\usepackage{multirow}
\usepackage{arydshln}
\input{preamble.tex}
\title{VisCoder: Fine-Tuning LLMs for Executable Python \\Visualization Code Generation}

\author{
 \textbf{Yuansheng Ni\textsuperscript{1}},
 \textbf{Ping Nie\textsuperscript{4}},
 \textbf{Kai Zou\textsuperscript{3}},
 \textbf{Xiang Yue\textsuperscript{2}},
 \textbf{Wenhu Chen\textsuperscript{1}},
\\
 \textsuperscript{1}University of Waterloo,
 \textsuperscript{2}Carnegie Mellon University,
 \textsuperscript{3}Netmind.ai,
 \textsuperscript{4}Independent Researcher,
\\
 \small{
   \texttt{\{yuansheng.ni, wenhuchen\}@uwaterloo.ca}
 }\\
 \url{https://tiger-ai-lab.github.io/VisCoder}
}

\begin{document}
\maketitle
\begin{abstract}
Large language models (LLMs) often struggle with visualization tasks like plotting diagrams, charts, where success depends on both code correctness and visual semantics. Existing instruction-tuning datasets lack execution-grounded supervision and offer limited support for iterative code correction, resulting in fragile and unreliable plot generation. We present \textbf{VisCode-200K}, a large-scale instruction tuning dataset for Python-based visualization and self-correction. It contains over 200K examples from two sources: (1) validated plotting code from open-source repositories, paired with natural language instructions and rendered plots; and (2) 45K multi-turn correction dialogues from Code-Feedback, enabling models to revise faulty code using runtime feedback. We fine-tune Qwen2.5-Coder-Instruct on VisCode-200K to create \textbf{VisCoder}, and evaluate it on PandasPlotBench. VisCoder significantly outperforms strong open-source baselines and approaches the performance of proprietary models like GPT-4o-mini. We further adopt a \textit{self-debug} evaluation protocol to assess iterative repair, demonstrating the benefits of feedback-driven learning for executable, visually accurate code generation.
\end{abstract}

\input{sec/1_introduction}
\input{sec/2_related_work}
\input{sec/3_data}
\input{sec/4_experiment}
\input{sec/5_results}

\input{sec/6_conclusion}

\clearpage
\section*{Limitations}
Although VisCoder substantially improves visualization code generation, its scope is currently limited to Python, leaving visualization tasks involving other programming languages such as R and JavaScript unexplored. Even within Python, performance on Plotly remains comparatively weaker due to its verbose syntax and complex API structure, frequently causing semantic execution errors that the existing self-debugging routine struggles to address. Furthermore, our evaluation relies on the default automatic judge model adopted from prior studies, without an independent analysis of its potential biases or reliability.


\bibliography{custom}

\input{sec/X_appendix}
\end{document}

%% file: preamble.tex
\usepackage{wrapfig}
\usepackage{xcolor}         
\usepackage{multirow}
\usepackage{graphicx}
\usepackage{subcaption}
\usepackage{enumitem}
\usepackage{tcolorbox}
\usepackage{color-edits}
\usepackage{tocloft}
\usepackage{booktabs}
\usepackage{adjustbox}
\usepackage{amsmath}
\newlistof{appfigures}{afg}{List of Comparision Figures in Different Settings}

\extrafloats{100}

\usepackage[normalem]{ulem}
\useunder{\uline}{\ul}{}

\addauthor{gn}{magenta}
\addauthor{ysu}{cyan}
\definecolor{darkblue}{RGB}{84, 112, 198}
\definecolor{lightgreen}{RGB}{145, 204, 117}
\definecolor{lightyellow}{RGB}{250, 200, 88}
\definecolor{lightred}{RGB}{238, 102, 102}
\definecolor{lightblue}{RGB}{115, 192, 222}

\newtcolorbox{promptbox}[2][Prompt]{
colback=black!5!white,
arc=5pt, 
boxrule=0.5pt,
fonttitle=\bfseries,
title=#1, 
before upper={\small}, fontupper=\fontfamily{ptm}\selectfont,
colframe=#2,
}

\usepackage{etoc}
\usepackage{titletoc}

\newcommand\DoToC{%
  \startcontents
  \printcontents{}{1}{\noindent \textbf{\Large{Table of Contents in Appendix}}\vskip3pt\vskip5pt}
  \vskip3pt\vskip5pt
}

\usepackage{pifont}

\newcommand{\greencheck}{\textcolor{green!60!black}{\ding{52}}} 
\newcommand{\redmark}{\textcolor{red}{\ding{55}}}

%% file: sec/1_introduction.tex
\section{Introduction}

Despite the growing capabilities of large language models (LLMs) in general-purpose code generation~\cite{codex,deepseek-coder}, they continue to struggle with one of the most common and visually essential tasks in data analysis: \textit{generating code that produces a valid and semantically meaningful plot}. For example, given a tabular description, models may generate code that appears syntactically correct and invokes the appropriate libraries~\cite{dibia2023lida, xie2024waitgpt}. But when executed, the result is often broken: exceptions are raised, plots render blank or malformed, or the visual fails to reflect the intended semantics of the instruction~\cite{chen2024viseval,yang2024matplotagent, galimzyanov2024drawing}.

These failures are not incidental: they reflect structural challenges in visualization code generation. Unlike standard text-to-code tasks, visualization requires grounding across three modalities: \textit{natural language} (the user instruction), \textit{data structure} (the tabular or other data input), and \textit{visual output} (the rendered chart). Execution correctness is not binary; a script may run and still fail to convey the intended meaning. Visualization libraries such as \texttt{matplotlib}~\cite{hunter2007matplotlib}, \texttt{seaborn}~\cite{waskom2021seaborn}, and \texttt{plotly}~\cite{inc2015collaborative} further complicate the task, with API idiosyncrasies and intricate bindings between data, layout, and style. Current instruction-tuning datasets do not meet the demands of this setting. Most lack explicit visual grounding, do not enforce runtime validation, and provide little to no supervision for recovery from failure. As a result, even the most advanced open models like Qwen-Coder~\cite{qwen2.5} or DeepSeek-Coder~\cite{deepseek-coder} struggle with executable, semantically accurate visualization code, particularly when debugging is required~\cite{ zheng2024opencodeinterpreter}.

To address these gaps, we introduce \textbf{VisCode-200K}, a new instruction-tuning dataset for Python-based visualization code generation and multi-turn correction. VisCode-200K contains over 200K supervised examples derived from two complementary data sources: 1) \textbf{Executable visualization code}, extracted from open-source Python repositories and filtered across widely-used plotting libraries, including \texttt{matplotlib}, \texttt{seaborn} and others. All code samples are validated for runtime executability and paired with rendered plots. Natural language instructions are generated using LLMs conditioned on both the code and its output image~\cite{galimzyanov2024drawing}. 2) \textbf{Multi-turn revision dialogues}, drawn from the Code-Feedback dataset~\cite{zheng2024opencodeinterpreter}, which contains realistic interactions where models revise faulty Python code based on runtime errors and follow-up prompts. While not visualization-specific, these traces provide essential supervision for teaching models to debug and recover from execution failures. This dual-source construction enables training for both \textit{single-shot generation} and \textit{multi-round refinement}, allowing models to generate code initially and improve it iteratively through feedback.

To demonstrate the effectiveness of VisCode-200K, we fine-tune Qwen2.5-Coder-Instruct~\cite{hui2024qwen2} at both 3B and 7B scales to produce \textbf{VisCoder}, an open-source model tuned specifically for Python visualization tasks. We evaluate VisCoder on \textbf{PandasPlotBench}~\cite{galimzyanov2024drawing}, a benchmark that assesses executable code generation from natural language and data previews across three plotting libraries. We also introduce a \textbf{self-debug evaluation mode}, in which models are given multiple rounds to revise failed outputs based on execution traces, simulating a realistic developer-style correction loop.

Our experiments show that VisCoder substantially outperforms competitive open-source baselines. VisCoder-3B and 7B achieve average execution pass rate improvements of \textbf{19.6} and \textbf{14.5} points over Qwen2.5-Coder. Under self-debug mode, it reaches over \textbf{90\%} execution pass rate on \texttt{Matplotlib} and \texttt{Seaborn}. Compared to proprietary models, VisCoder-7B surpasses GPT-4o-mini on both \texttt{Seaborn} and \texttt{Plotly} under the default setting, and approaches GPT-4o performance on both libraries after self-debugging. At 3B scale, it outperforms GPT-4o-mini on \texttt{Seaborn} and narrows the gap in other libraries. These results demonstrate the impact of combining domain-specific instruction tuning with feedback-driven correction for grounded visualization code generation.

%% file: sec/2_related_work.tex
\section{Related Work}
\paragraph{LLMs for Visualization Code Generation.}

Recent work has explored using large language models to generate visualization code from natural language prompts. Benchmarks such as MatPlotAgent and VisEval~\cite{yang2024matplotagent, chen2024viseval} evaluate model performance on structured NL2VIS tasks with paired chart specifications and data previews, while PandasPlotBench~\cite{galimzyanov2024drawing} provides a curated benchmark for assessing executable visualization code generation across multiple plotting libraries. Plot2Code~\cite{wu2024plot2code} and ChartMimic~\cite{yang2024chartmimic} explore the reverse direction by generating code from visual inputs, using rendered plots or chart images with instructions, and shift focus from textual reasoning to cross-modal understanding. These studies highlight persistent challenges in semantic grounding, API correctness, and robustness across different plotting tasks. Broader evaluations have analyzed model behavior across visualization types and libraries~\cite{vázquez2024llmsreadyvisualization, podo2024vi}, while specification-based approaches using Vega-Lite~\cite{xie2024waitgpt} offer an alternate formulation that lacks direct executability. Beyond evaluation, systems like LIDA~\cite{dibia2023lida} and VisPath~\cite{seo2025vispath} incorporate summarization, code synthesis, and feedback-driven refinement into end-to-end pipelines. Related efforts have also extended visual code generation to structured domains such as parametric CAD modeling~\cite{li2025cad} and mathematical animation~\cite{ku2025theoremexplainagent}, where outputs reflect domain-specific constraints rather than general-purpose charting semantics. However, most prior work lacks training data grounded in execution outcomes and provides limited support for iterative refinement. These limitations hinder model reliability, especially when generating code that must be both syntactically correct and semantically faithful to the intended visualization.

\paragraph{Execution Feedback and Code Correction.}

Execution feedback has been widely explored as a supervisory signal for improving the reliability of code generation. Prior work investigates using runtime traces to guide post-hoc refinement~\cite{jain2025multi, chen2025revisit, tian2024debugbench, zhang2025extracting}, or integrates such signals into training through reinforcement learning~\cite{gehring2024rlef, zeng2025acecoder}. Other approaches emphasize multi-turn correction, where models revise faulty code using internal or external feedback~\cite{madaan2023self, jiang2024ledex, zheng2024opencodeinterpreter, ruiz2025art}, or simulate debugging workflows with planning and agent collaboration~\cite{grishina2025fully, li2024codetree}. In the context of visualization, VisPath~\cite{seo2025vispath} and MatPlotAgent~\cite{yang2024matplotagent} explore chart refinement using visual feedback from rendered outputs. Yet despite these advances, supervision grounded in execution feedback or revision traces has rarely been used to train models for visualization code generation, where runtime validity and semantic alignment remain central challenges.

%% file: sec/3_data.tex
\begin{figure*}[ht]
    \centering
    \includegraphics[width=1\linewidth]{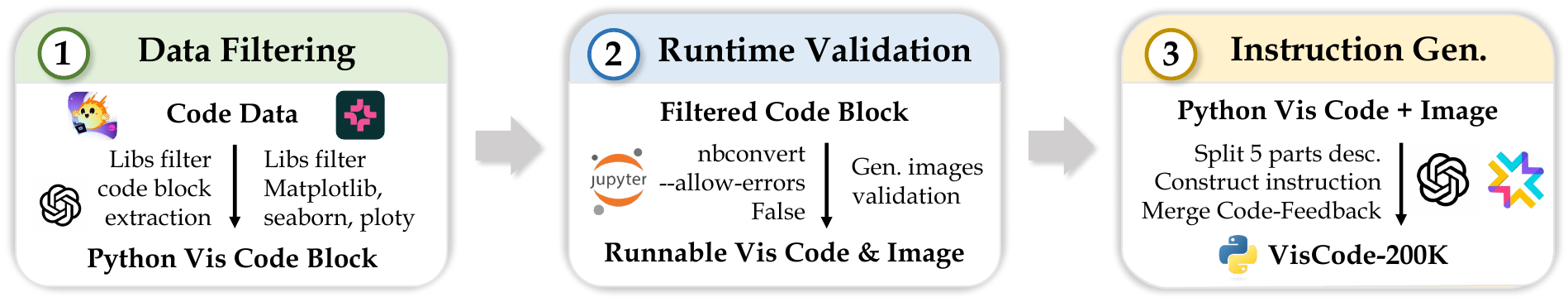}
    \caption{Data construction pipeline for VisCode-200K. We extract and filter visualization code blocks from open-source Python sources, validate their executability and plot rendering via Jupyter-based runtime checks, and generate structured instructions paired with rendered plots. We integrate multi-turn correction data from Code-Feedback during instruction construction to support iterative refinement.}
    \label{fig:data_pipeline}
\end{figure*}
\section{VisCode-200K: A Python Visualization Instruction Tuning Dataset}

In this section, we present \textbf{VisCode-200K}, a supervised instruction tuning dataset for Python-based visualization and feedback-driven code correction. It is designed to support robust code generation across diverse plotting libraries and to enable iterative refinement through multi-turn supervision.

VisCode-200K integrates two complementary sources of supervision. The first consists of executable visualization code extracted from open-source Python repositories, covering a wide range of real-world chart types, layouts, and plotting libraries. All samples are filtered to ensure runtime validity and compatibility with standard Python environments, exposing models to diverse and realistic plotting practices. The second source comprises multi-turn Python dialogues from the Code-Feedback dataset~\citep{zheng2024opencodeinterpreter}, which offer supervision for revising faulty code in response to execution errors. While not specific to visualization, these interactions are critical for modeling realistic correction behaviors in iterative workflows.

~\autoref{fig:data_pipeline} provides an overview of the VisCode-200K construction pipeline, which consists of code filtering, runtime validation, and structured instruction generation. The following subsections detail each component.

\subsection{Code Extraction from Public Repositories}

To build a large corpus of executable Python visualization code, we source data from two open datasets: the Python subset of \texttt{stack-edu}\footnote{\href{https://huggingface.co/datasets/HuggingFaceTB/stack-edu}{hf.co/datasets/HuggingFaceTB/stack-edu}}~\citep{allal2025smollm2smolgoesbig} and the chart/table partitions of \texttt{CoSyn-400K}\footnote{\href{https://huggingface.co/datasets/allenai/CoSyn-400K}{hf.co/datasets/allenai/CoSyn-400K}}~\citep{yang2025scaling, deitke2024molmo}. From these corpora, we extract code that uses commonly adopted visualization libraries, including \texttt{matplotlib}, \texttt{seaborn} and others, to ensure broad coverage of real-world plotting styles. The construction pipeline consists of four stages: library-based filtering, code block extraction, runtime validation, and instruction generation.

\paragraph{Filtering and Code Block Extraction.}  
For the \texttt{stack-edu} source, which contains a large collection of Python code examples from educational contexts, we begin by applying library-based filters to identify approximately 1.7M samples that invoke common Python visualization libs. Since most examples embed visualization logic within broader program contexts, we use GPT-4o-mini~\citep{gpt4o-mini} to extract minimal, standalone plotting blocks. During this process, we inject mock data to replace missing inputs and ensure that each block can be executed in isolation. This structural cleaning step yields code samples that reflect realistic plotting usage while remaining compatible with our runtime pipeline. After filtering and reconstruction, we obtain roughly 1M candidate blocks. To balance library distribution, we retain all \texttt{seaborn} and ohter samples and randomly subsample a matching number of \texttt{matplotlib} examples, resulting in a curated subset of ~300K visualization blocks.

From \texttt{CoSyn-400K}, we extract 112K Python code snippets that include calls to one of the target visualization libraries. CoSyn provides high-quality synthetic plotting code spanning a wide range of styles, with well-rendered outputs and consistent structure. Unlike \texttt{stack-edu}, it stores code and data separately, which requires reconstruction to enable runtime execution. We synthesize runnable scripts by inserting inline annotations such as column headers and the first data row to emulate realistic \texttt{pandas.read\_csv} loading. When necessary, we append missing plotting function calls to ensure that each script can execute fully within a notebook environment.

\paragraph{Runtime Validation.}
To verify executability, we run each code block in an isolated Jupyter environment using \texttt{nbconvert} with \texttt{allow-error=False}. We enforce a timeout and terminate executions that hang or enter infinite loops using a simulated keyboard interrupt. Only samples that run successfully and generate a valid image file are retained. This step yields 105K validated plotting scripts from \texttt{stack-edu} and 50K from \texttt{CoSyn-400K}, each paired with its corresponding output image.

\paragraph{Instruction Generation.}

To construct meaningful instructions for visualization code generation, we use GPT-4o~\citep{gpt4o} to synthesize instruction components based on each validated code block and its corresponding plot. This enables the model to incorporate both structural code features and visual semantics from the rendered image.

Each instruction consists of five components: (1) a brief setup description specifying the programming language and visualization libraries used; (2) a data description summarizing the tabular input and column semantics; (3) a data block indicating the input table, either as mock data (for \texttt{stack-edu}) or a two-row preview (for \texttt{CoSyn}); (4) a high-level plot description outlining axes and structural layout; and (5) a style description capturing colors, grid layout, and other visual properties.

For \texttt{stack-edu} samples, mock data is extracted directly from the code block, where it was inserted during preprocessing. For \texttt{CoSyn}, where data is stored separately, we construct a compact preview using the first two rows of the table. The five components are then assembled using a fixed template to form the final instruction:

\vspace{-2pt}
\begin{quote}
\small
\texttt{[Plot Description]} \\
\texttt{[Setup]} \\
\texttt{[Data Description]} \\
\texttt{"The mock data shows below:" or "The first two rows of the data are shown below:"} \\
\texttt{[Data]} \\
\texttt{[Plot Style Description]}
\end{quote}
\vspace{-2pt}

This format enforces a consistent prompt structure across both data sources, providing models with a clear description of the target plot as well as the data and style required to render it.

\subsection{Multi-turn Instruction-following Dialogues with Execution Feedback}

To train models with self-correction capabilities, we incorporate 45K multi-turn dialogues from the \texttt{Code-Feedback}\footnote{\href{https://huggingface.co/datasets/m-a-p/Code-Feedback}{hf.co/datasets/m-a-p/Code-Feedback}} dataset~\citep{zheng2024opencodeinterpreter}. These dialogues involve Python-based tasks, including user instructions, model-generated code, and follow-up turns containing execution feedback or revision prompts.

We begin with 56K Python dialogues and remove those with excessive length or turn count to maintain consistency and reduce training complexity. The resulting 45K samples span diverse Python tasks with realistic correction behaviors.

While not specific to visualization, these dialogues offer valuable supervision for teaching models to revise faulty code based on runtime signals and to reason over iterative interactions. We integrate them into the instruction tuning corpus alongside the single-turn samples from \texttt{stack-edu} and \texttt{CoSyn}, enabling models to learn both initial generation and multi-turn refinement strategies.

%% file: sec/4_experiment.tex
\section{Experiment Setup}
\paragraph{Training Setup.}
We fine-tune Qwen2.5-Coder-Instruct~\citep{hui2024qwen2} at two parameter scales: 3B and 7B. This allows us to assess the generalizability of VisCode-200K across different model capacities. Both models are trained for 3 epochs with a learning rate of $5 \times 10^{-6}$, a warm-up ratio of 0.05, and a cosine learning rate scheduler. We perform full-parameter tuning in bfloat16 precision on 8$\times$A100 GPUs with a total batch size of 128, using SWIFT infrastructure~\citep{zhao2024swiftascalablelightweightinfrastructure}.

\paragraph{Evaluation Setup.}
We evaluate models using PandasPlotBench~\citep{galimzyanov2024drawing}, a benchmark designed to assess the ability of language models to generate executable and semantically accurate visualization code from tabular data descriptions. It contains 175 tasks spanning three widely used Python plotting libraries: \texttt{matplotlib}, \texttt{seaborn}, and \texttt{plotly}.

Each task includes a natural language instruction and a preview of the input DataFrame. The model is expected to generate Python code that produces a valid plot when executed according to the instruction. The benchmark reports three metrics: (1) \textit{Incorrect Code Rate}, the proportion of outputs that fail to produce any plot; and two GPT-4o-judged scores: (2) a task-based score measuring alignment with the instruction, and (3) a visual score assessing similarity to the reference plot.

Among these metrics, Incorrect Code Rate provides only a coarse signal of success. It indicates whether a plot is rendered, but does not capture execution errors if a figure is produced. As a result, blank or semantically meaningless outputs—such as plots with only axes—may be misclassified as correct. To address this issue, we introduce an additional metric: \textbf{Execution Pass Rate}, defined as the percentage of outputs that execute without error.

\paragraph{Self-Debug Evaluation Mode.}
To evaluate a model’s ability to recover from failure, we extend the benchmark with a \textit{self-debug evaluation mode}. In this setting, if the initial generation fails to execute or does not produce a valid plot, the model is allowed up to $K$ rounds to iteratively revise its output based on accumulated feedback.

At each round, only the tasks that remain unsolved from the previous attempt are reconsidered. The model receives a multi-turn prompt constructed as a dialogue, including the original instruction, its failed code response, and a follow-up message requesting correction based on the execution error. Conditioned on this dialogue history, the model generates a revised version of the code. Tasks are considered successfully fixed if the generated code executes without error and produces a valid plot. These tasks are excluded from subsequent rounds.

\vspace{-3pt}

\begin{algorithm}[!h]
\caption{Self-Debug Evaluation Protocol}
\label{alg:self-debug}
\begin{algorithmic}[1]
\STATE Let $F_0$ be failed tasks from initial evaluation
\FOR{$i = 1$ to $K$}
    \FOR{each task $x$ in $F_{i-1}$ not yet fixed}
        \STATE Fix $x$ via feedback-driven prompting
        \STATE Evaluate the result of the revised code
        \IF{successful}
            \STATE Mark $x$ as fixed \& record output
        \ELSE
            \STATE Record $x$'s latest failed output
        \ENDIF
    \ENDFOR
\ENDFOR
\STATE Evaluate all tasks with final recorded outputs
\end{algorithmic}
\end{algorithm}

\vspace{-3pt}

We set $K = 3$ for all experiments. After the final round of self-debug, each task is evaluated based on its recorded final output, which is either the successfully revised version from an earlier round or the last failed attempt if no fix was found. The resulting outputs are scored using the same evaluation pipeline as in the default setting. The full procedure is summarized in Algorithm~\ref{alg:self-debug}.

This iterative process simulates a developer-style debugging loop and enables systematic evaluation of the model’s ability to recover from failure through multi-round code correction.

%% file: sec/5_results.tex
\begin{table*}[t]
\centering
\resizebox{\linewidth}{!}{
\begin{tabular}{lccccccccccccccc}
\toprule
\multirow{3}{*}{\textbf{Model}} & \multicolumn{5}{c}{\textbf{Matplotlib}} & \multicolumn{5}{c}{\textbf{Seaborn}} & \multicolumn{5}{c}{\textbf{Plotly}} \\
\noalign{\vskip 2pt}
 & \multicolumn{1}{l}{\multirow{2}{*}{
 \textbf{\begin{tabular}[c]{@{}l@{}}Exec\\ Pass\end{tabular}}}
 } & \multicolumn{2}{c}{\textbf{Mean}} & \multicolumn{2}{c}{\textbf{Good($\geq$75)}} & \multicolumn{1}{l}{\multirow{2}{*}{
 \textbf{\begin{tabular}[c]{@{}l@{}}Exec\\ Pass\end{tabular}}}
 } & \multicolumn{2}{c}{\textbf{Mean}} & \multicolumn{2}{c}{\textbf{Good($\geq$75)}} & \multicolumn{1}{l}{\multirow{2}{*}{
 \textbf{\begin{tabular}[c]{@{}l@{}}Exec\\ Pass\end{tabular}}}
 } & \multicolumn{2}{c}{\textbf{Mean}} & \multicolumn{2}{c}{\textbf{Good($\geq$75)}} \\
&  & vis & task & vis & task &  & vis & task & vis & task &  & vis & task & vis & task \\
\midrule
GPT-4o & 94.9 & 75 & 90 & 67\% & 93\% & {\ul 83.4} & 65 & 78 & 59\% & 80\% & {\ul 77.7} & 55 & 68 & 50\% & 70\% \\
GPT-4o + Self Debug & \textbf{99.4} & 77 & 93 & 69\% & 96\% & \textbf{92.6} & 69 & 84 & 63\% & 86\% & \textbf{97.7} & 68 & 84 & 61\% & 83\% \\
GPT-4o-mini & 88.6 & 68 & 86 & 59\% & 86\% & 62.3 & 45 & 57 & 41\% & 57\% & 69.1 & 48 & 52 & 42\% & 51\% \\
GPT-4o-mini + Self Debug & {\ul 97.7} & 72 & 92 & 65\% & 94\% & 72.0 & 47 & 60 & 43\% & 61\% & \textbf{97.7} & 62 & 71 & 51\% & 67\% \\ \midrule
\multicolumn{16}{c}{$\sim$ 3B Scale} \\ \midrule
Llama-3.2-3B-Ins. & 65.1 & 43 & 60 & 34\% & 55\% & 30.9 & 18 & 24 & 14\% & 21\% & 13.1 & 8 & 8 & 7\% & 8\% \\
Qwen-2.5-3B-Ins. & 74.3 & 55 & 68 & 49\% & 66\% & 58.3 & 43 & 58 & 33\% & 51\% & 30.9 & 19 & 23 & 17\% & 21\% \\
Qwen-2.5-Coder-3B-Ins. & 71.4 & 56 & 72 & 50\% & 69\% & 58.3 & 44 & 55 & 36\% & 51\% & 27.4 & 17 & 19 & 17\% & 18\% \\
\noalign{\vskip 2pt}
\hdashline
\noalign{\vskip 2pt}
\textbf{VisCoder-3B} & {\ul 81.7} & 60 & 69 & 53\% & 69\% & {\ul 73.7} & 48 & 65 & 38\% & 61\% & {\ul 60.6} & 38 & 45 & 32\% & 44\% \\
\textbf{VisCoder-3B + Self Debug} & \textbf{85.1} & 60 & 70 & 53\% & 69\% & \textbf{78.3} & 48 & 66 & 37\% & 62\% & \textbf{64.6} & 40 & 48 & 34\% & 47\% \\ \midrule
\multicolumn{16}{c}{$\sim$ 7B Scale} \\ \midrule
Llama-3.1-8B-Ins. & 81.1 & 61 & 76 & 51\% & 74\% & 65.7 & 51 & 64 & 45\% & 63\% & 30.9 & 21 & 22 & 20\% & 21\% \\
Qwen2.5-7B-Ins. & 77.1 & 64 & 76 & 53\% & 75\% & 66.3 & 51 & 63 & 46\% & 62\% & 56.0 & 38 & 42 & 31\% & 40\% \\
Qwen2.5-Coder-7B-Ins. & 78.3 & 63 & 76 & 58\% & 75\% & 68.6 & 51 & 63 & 40\% & 62\% & 48.0 & 29 & 34 & 24\% & 31\% \\
\noalign{\vskip 2pt}
\hdashline
\noalign{\vskip 2pt}
\textbf{VisCoder-7B} & {\ul 87.4} & 66 & 78 & 60\% & 80\% & {\ul 76.6} & 57 & 70 & 50\% & 68\% & {\ul 74.3} & 48 & 60 & 41\% & 61\% \\
\textbf{VisCoder-7B + Self Debug} & \textbf{91.4} & 67 & 81 & 62\% & 83\% & \textbf{90.3} & 62 & 77 & 51\% & 75\% & \textbf{81.7} & 51 & 65 & 44\% & 65\% \\
\bottomrule
\end{tabular}
}
\caption{Performance of selected models on the PandasPlotBench benchmark. For each model, we report (1) execution pass rate (\textbf{Exec Pass}), (2) mean visual and task scores (\textbf{Mean}), and (3) the proportion of samples scoring at least 75 (\textbf{Good}). The best-performing model in each scale is shown in \textbf{bold}, and the second best is {\ul{underlined}}.}
\label{tab:overall_results}
\end{table*}

\section{Main Results}
We present the main experimental results on PandasPlotBench, including overall model comparisons, performance under the self-debug evaluation protocol, error type analysis, and a training data ablation study. More detailed results are provided in ~\autoref{sec:extended_results}.

\subsection{Overall Model Comparison}

We evaluate VisCoder models against both proprietary and open-source language models to assess executable visualization performance across scales and libraries. The proprietary group includes GPT-4o~\citep{gpt4o}, the strongest model in the original PandasPlotBench benchmark, and its lightweight variant GPT-4o-mini~\citep{gpt4o-mini}. Among open-source baselines, we compare LLaMA-3.2-3B, LLaMA-3.1-8B~\citep{grattafiori2024llama}, Qwen2.5-Instruct, and Qwen2.5-Coder-Instruct~\citep{qwen2.5, hui2024qwen2}, evaluated at both 3B and 7B scales. VisCoder models are trained on VisCode-200K and fine-tuned using the same instruction tuning setup.

~\autoref{tab:overall_results} summarizes model performance across the three plotting libraries. The following analysis focuses on execution success, task alignment, and visual fidelity, highlighting VisCoder’s comparative strengths and remaining challenges.

\paragraph{Proprietary Models Remain Stronger.}
Proprietary models outperform open-source models by a wide margin across all plotting libraries. GPT-4o achieves the highest execution pass rates and the strongest judge-based scores, followed by its lightweight variant GPT-4o-mini. These results indicate more reliable execution and better semantic alignment with task instructions, particularly in complex visualization settings. In contrast, open-source models such as LLaMA and Qwen2.5-Instruct consistently underperform across all metrics. This reinforces the gap between proprietary and open-source systems on execution-sensitive and semantically grounded code generation.

\paragraph{Plotly Presents Harder Challenge.}
Performance differs across plotting libraries. While most models perform reliably on \texttt{matplotlib} and \texttt{seaborn}, results on \texttt{plotly} are markedly lower, especially for open-source models. Execution pass rates often fall below 35\%, and task and visual scores drop accordingly. Generated plots frequently fail to reflect the intended semantics or produce complete visuals. This suggests that \texttt{plotly}'s verbose syntax and less represented API structure pose greater challenges for current models.

\paragraph{VisCoder Closes the Open-Source Gap.}
VisCoder models consistently outperform their untuned Qwen2.5-Coder baselines across all libraries. At 3B, VisCoder improves both execution success and semantic alignment, with larger gains on \texttt{plotly} and \texttt{seaborn}, where baseline generations often fail to capture visual intent. At 7B, VisCoder outperforms GPT-4o-mini on both \texttt{seaborn} and \texttt{plotly}, while remaining slightly behind on \texttt{matplotlib}. These results demonstrate that domain-specific instruction tuning improves functional reliability and output fidelity, especially in libraries with more complex plotting structures.

\paragraph{Self-Debug Further Boosts Performance.}
GPT-4o demonstrates strong self-debugging ability, reaching near-perfect execution pass rates after multiple rounds of correction. VisCoder models also improve substantially under this protocol. VisCoder-7B surpasses 90\% execution success on both \texttt{matplotlib} and \texttt{seaborn}, with especially large gains on the latter. Task and visual scores improve consistently across rounds. These results show that VisCoder can generalize from its training data to refine failed outputs over multiple attempts, even without task-specific debugging supervision.

\subsection{Self-Debug Evaluation Results}

\begin{figure*}[ht]
    \centering
    \includegraphics[width=1\linewidth]{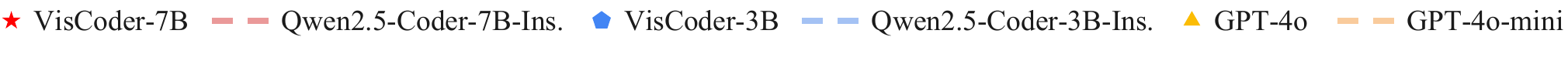}
    \begin{subfigure}[t]{0.32\textwidth}
        \includegraphics[width=\linewidth]{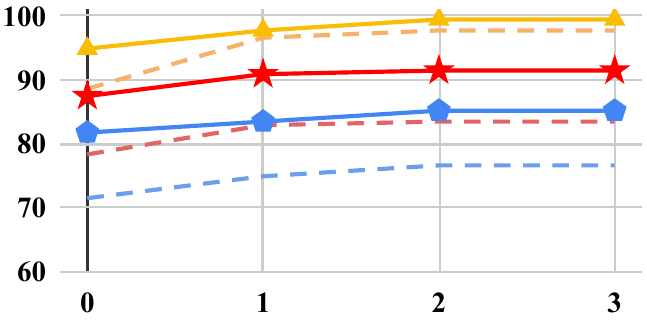}
        \caption{Matplotlib}
        \label{fig: slef_debug_matplotlib}
    \end{subfigure}
    \hfill
    \begin{subfigure}[t]{0.32\textwidth}
        \includegraphics[width=\linewidth]{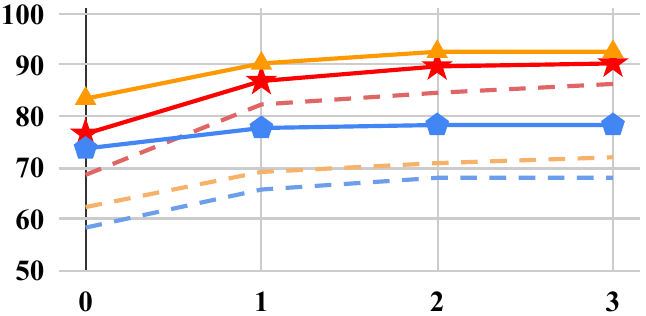}
        \caption{Seaborn}
        \label{fig: slef_debug_seaborn}
    \end{subfigure}
    \hfill
    \begin{subfigure}[t]{0.32\textwidth}
        \includegraphics[width=\linewidth]{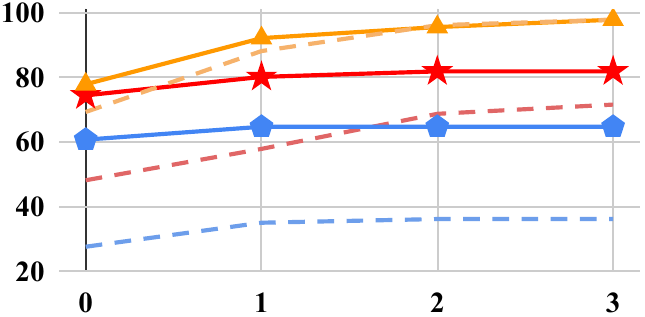}
        \caption{Plotly}
        \label{fig: slef_debug_plotly}
    \end{subfigure}
    \caption{\textbf{Execution pass rate} across self-debug rounds (Attempt 0–3), shown separately for three plotting libraries. Attempt 0 corresponds to the default output, while Attempts 1–3 represent subsequent correction rounds. Model groups are color-coded, with solid and dashed lines used to distinguish paired models. VisCoder models improve consistently across rounds, with VisCoder-7B gradually closing the gap to GPT-4o on \texttt{seaborn}. Y-axis ranges are scaled per subplot to match library-specific score distributions.}
    \label{fig:self_debug_lineplot}
\end{figure*}

To analyze the dynamics of self-debugging, we track execution pass rates over multiple correction rounds by evaluating GPT-4o and GPT-4o-mini as proprietary baselines, alongside VisCoder models at 3B and 7B scales. To isolate the effects of instruction tuning, we also include untuned Qwen2.5-Coder models at matching sizes. ~\autoref{fig:self_debug_lineplot} shows execution pass rates from the initial generation (Attempt 0) through three rounds of self-debugging (Attempts 1–3), presented separately for each plotting library. Detailed breakdown of pass rates per model and library is provided in ~\autoref{sec: breakdown_self_debug_results}.

\paragraph{Self-debug is broadly effective.}
Execution pass rates increase steadily over self-debug rounds for most models and libraries, indicating the overall effectiveness of the protocol. The first attempt typically yields the largest improvement, with smaller gains in subsequent rounds. This pattern suggests that a simple retry mechanism informed by execution feedback can recover a substantial portion of initial failures.

\paragraph{VisCoder yields stable behavior.}
Compared to their Qwen2.5-Coder baselines, VisCoder models show smaller per-round gains in execution pass rate but consistently achieve higher final performance. This suggests that VisCoder tends to generate stronger initial outputs and applies more stable corrections across rounds. The effect is most pronounced with VisCoder-7B on \texttt{seaborn}, where execution rates increase steadily and approach GPT-4o by the final attempt.

\paragraph{Failures remain across models.}
Even the strongest model GPT-4o does not reach perfect execution rates after self-debugging. On \texttt{seaborn}, its performance plateaus after three rounds, leaving a non-trivial portion of failures unresolved. In contrast, VisCoder-3B stands out among small-scale models. It surpasses GPT-4o-mini on \texttt{seaborn} and performs competitively across other libraries. Meanwhile, we observe that smaller models tend to reach their performance ceiling more quickly, exhibiting smoother but more limited improvements across rounds.

\subsection{Error Analysis}

To examine the error recovery behavior of VisCoder-7B, we analyze how execution error counts transition before and after self-debugging. \autoref{tab:self_debug_errors} summarizes four representative error types, grouped by plotting library. A detailed breakdown by model and debug round is provided in Appendix~\ref{sec: breakdown_self_error_analysis}.

\begin{table}[ht]
\centering
\resizebox{0.93\linewidth}{!}{
\begin{tabular}{lccc}
\toprule
\noalign{\vskip 1pt}
\textbf{Error Type} & \textbf{Matplotlib} & \textbf{Seaborn} & \textbf{Plotly} \\
\noalign{\vskip 1pt}
\midrule
AttributeError & 5 $\rightarrow$ 2 & 15 $\rightarrow$ 2 & 5 $\rightarrow$ 1 \\
TypeError      & 7 $\rightarrow$ 5 & 8 $\rightarrow$ 4  & 3 $\rightarrow$ 1 \\
\noalign{\vskip 2pt}
\hdashline
\noalign{\vskip 2pt}
KeyError       & 1 $\rightarrow$ 1  & 0 $\rightarrow$ 0   & 1 $\rightarrow$ 1 \\
ValueError     & 4 $\rightarrow$ 5 & 8 $\rightarrow$ 7  & 29 $\rightarrow$ 23 \\
\bottomrule
\end{tabular}
}
\caption{Execution error count transitions for VisCoder-7B across four representative error types, segmented by plotting library. Each value shows the transition from the initial to the post-debugging error count (X → Y).}
\label{tab:self_debug_errors}
\end{table}

\paragraph{Effective Recovery from Structural Errors.}

VisCoder-7B demonstrates strong self-correction ability on shallow, structural errors. \textit{AttributeErrors} in Seaborn are reduced from 15 to 2, and \textit{TypeErrors} in Plotly from 3 to 1. These failures typically result from incorrect method calls, invalid argument types, or simple syntax mistakes, and are often accompanied by clear diagnostic messages. As illustrated in \autoref{fig:case_matplotlib_debug} and \autoref{fig:case_seaborn_debug}, VisCoder can reliably correct such cases using runtime feedback, frequently producing valid plots on retry.

\paragraph{Persistent Failures in Semantic Execution Errors.}
Semantic execution errors such as \textit{KeyError} and \textit{ValueError} remain difficult to resolve (\autoref{fig:case_plotly_debug}). On Plotly, \textit{ValueErrors} decrease from 29 to 23 across three rounds of correction, but a substantial number still remain. Meanwhile, \textit{KeyErrors} show no improvement, remaining at 1 throughout. These failures are often caused by invalid trace configurations or mismatched array lengths and typically require reasoning over the input data structure. However, the model does not dynamically reassess the DataFrame during self-debug, leading to retries that rely on faulty assumptions. Compared to structural errors, semantic failures are less localized and more difficult to resolve through symbolic correction alone.

\subsection{Training Data Ablation}

We assess the contribution of each training data source in VisCode-200K through a controlled ablation study, including two reference points: the model trained on the full VisCode-200K dataset and the untuned Qwen2.5-Coder-7B-Instruct baseline. Separate Qwen2.5-Coder-7B models are fine-tuned on subsets from \texttt{stack-edu}, \texttt{CoSyn-400K}, and \texttt{Code-Feedback}, using the same instruction tuning setup as the full configuration. All models are evaluated on PandasPlotBench under both default and self-debug modes. ~\autoref{tab: ablation} shows execution pass rates across the three plotting libraries.

\begin{table}[ht]
\resizebox{\linewidth}{!}{
\begin{tabular}{lcccc}
\toprule
\textbf{Model} & \textbf{Self-Debug} & \textbf{Matplotlib} & \textbf{Seaborn} & \textbf{Plotly} \\ \midrule
\multirow{2}{*}{Qwen2.5-Coder-7B-Ins} & \redmark & 78.3 & 68.6 & 48.0 \\
 & \greencheck & 83.4 & 86.3 & 71.4 \\
 \noalign{\vskip 2pt}
\hdashline
\noalign{\vskip 2pt}
\multirow{2}{*}{+ Stack-Edu-105K} & \redmark & 66.3 & 55.4 & 49.7 \\
 & \greencheck & 72.0 & 69.7 & 61.1 \\
 \noalign{\vskip 2pt}
\hdashline
\noalign{\vskip 2pt}
\multirow{2}{*}{+ CoSyn-50K*} & \redmark & 0.0 & 0.0 & 5.7 \\
 & \greencheck & 0.0 & 0.0 & 6.3 \\
 \noalign{\vskip 2pt}
\hdashline
\noalign{\vskip 2pt}
\multirow{2}{*}{+ Code-Feedback-45K} & \redmark & 88.0 & 44.0 & 62.9 \\
 & \greencheck & 90.9 & 59.4 & 77.7 \\
 \noalign{\vskip 2pt}
\hdashline
\noalign{\vskip 2pt}
\multirow{2}{*}{+ VisCode-200K} & \redmark & 87.4 & 76.6 & 74.3 \\
 & \greencheck & 91.4 & 90.3 & 81.7 \\
 \bottomrule
\end{tabular}
}
\caption{Execution pass rates of Qwen2.5-Coder-7B models trained on individual subsets of VisCode-200K. Each model is evaluated across three libraries under both default (\redmark) and self-debug (\greencheck) modes.}
\label{tab: ablation}
\end{table}

\vspace{-6pt}

\paragraph{Stack-Edu provides moderate generalization.}
Using the subset from \texttt{stack-edu} results in modest gains over the baseline in \texttt{plotly} under the default setting (+1.7), but leads to significant drops on \texttt{matplotlib} and \texttt{seaborn} (–12.0 and –13.2). Self-debug improves pass rates across all libraries compared to their respective defaults, yet all scores remain below the untuned baseline. These results suggest that while stack-edu offers broad task coverage, it lacks the structural supervision and feedback-guided correction patterns needed for robust generalization.

\paragraph{CoSyn fails to generalize.}
The subset from \texttt{CoSyn-400K} fails to support effective instruction tuning for this task. Execution pass rates remain near zero across all libraries, and self-debug yields no meaningful improvement. Generated outputs often exhibit decoding instability, including repeated sequences, empty completions, or irrelevant boilerplate. A key reason is the homogeneous structure of the source data: all samples follow a fixed format consisting of imports, function definitions, and single function calls, which severely limits structural diversity during training. Combined with the synthetic and non-executable nature of the examples, this makes the single CoSyn subset ill-suited for executable visualization code generation.

\paragraph{Code-Feedback enhances structure but lacks breadth.}
The subset from \texttt{Code-Feedback} improves execution reliability on \texttt{matplotlib} and \texttt{plotly} in the default setting, outperforming the baseline by 9.7 and 14.9 points, respectively. These gains suggest that examples grounded in execution feedback help the model generate structurally valid and complete code. However, performance on \texttt{seaborn} remains low (44.0), and gains on \texttt{plotly} are limited compared to the full model. This reflects the general-purpose nature of the source data, which is not designed for visualization and lacks the task-specific grounding needed for broader transfer. Self-debug improves pass rates across libraries, but overall performance remains below that achieved with our full VisCode-200K dataset.

\paragraph{Full data offers complementary gains.}
The full VisCode-200K dataset yields the most consistent execution improvements across all plotting libraries and evaluation modes. Its performance under self-debug is particularly robust, with high pass rates maintained across structurally diverse tasks. These results reinforce the importance of domain-specific instruction tuning and multi-turn correction data for building reliable visualization-capable models.

%% file: sec/6_conclusion.tex
\section{Conclusion}

In conclusion, VisCode-200K provides a large-scale instruction tuning dataset for Python visualization code generation, combining executable plotting examples with multi-turn correction dialogues grounded in runtime feedback. To validate its effectiveness, we evaluate VisCoder models on PandasPlotBench using the default setting. Additionally, 
We propose a self-debug protocol to simulate realistic correction workflows and assess model performance in this extended evaluation mode.

Experiments show that VisCoder substantially outperforms strong open-source baselines across execution and alignment metrics, and narrows the gap to proprietary models like GPT-4o-mini. Gains are particularly pronounced in settings that involve complex visualization structures, such as Plotly, and iterative correction through self-debugging. Ablation studies further demonstrate that structurally diverse, executable training data and feedback-driven supervision contribute to more robust performance across plotting libraries.

Looking forward, this work reinforces the importance of domain-specific instruction tuning and multi-turn correction supervision for building robust and semantically grounded visualization-capable models. Future extensions may explore broader plotting libraries, richer correction supervision, and evaluation methods that measure models' abilities to recover from execution errors.

%% file: sec/X_appendix.tex
\appendix
\newpage
\clearpage
\onecolumn
\phantomsection
\label{list:list_of_appendix}
\DoToC
\clearpage
\input{appendix/prompt_template}

\input{appendix/self_debug_results}
\input{appendix/error_analysis_results}
\input{appendix/supp_results}
\input{appendix/case_study}

%% file: appendix/prompt_template.tex
\newpage
\section{Prompts Used for Dataset Construction}
In this section, we present the system prompts used during the construction of VisCode-200K. These prompts guide the automatic extraction of standalone visualization code from mixed-context sources, and support the generation of structured natural language instructions aligned with rendered plots.

\begin{promptbox}[Code Extraction Prompt]{darkblue}
\textbf{Model: GPT-4o-mini}\\ \\
You are a Python code extraction agent.\\

Given a Python code snippet and the used library, your task is to extract a self-contained and \textbf{runnable Python code block} that demonstrates how the specified library is actually used in the original code.

Use mock data where needed (e.g., pandas DataFrame, NumPy arrays), but keep it minimal and logically aligned with the original usage.  Retain any important structure, function calls, or plotting styles that reflect meaningful usage of the library.

- Do not include `plt.close()` or similar calls.\\
- If the library is only imported but never used, or if there is insufficient information to construct a meaningful runnable code block, return "null" (a literal string).\\
- Return only the Python code block enclosed in triple backticks like this: ```python ... ```, with nothing else.
\\ \\
Used Library:
\texttt{\{used\_libs\}}

Code:
\texttt{\{code\}}

\end{promptbox}

\begin{promptbox}[Instruction Generation Prompt: \texttt{stack-edu}]{gray}
\textbf{Model: GPT-4o}\\ 

Write the general TASK to write a code for plotting the given mock data. \\ \\
The code with mock data is given below, and the result of the generated plot image is given at the end. \\ \\
Split task into five parts:\\
1. Setup (describe programming language and libraries required to generate the plot).\\
2. Data Description (some short description of the mock data).\\
3. Data Generation (the data‑generation lines copied verbatim).\\
4. Plot Description (describe the structural layout of the plot, without referencing libraries or function names. Begin with “Generate...” or “Create...”).\\
5. Plot Style Description (describe the visual styling aspects of the plot, without referencing libraries or function).\\ \\
CODE:
\texttt{\{code\}} \\ \\
Each part of the task must start on a new line, numbered 1 through 5. Use plain text only. Do not include any markdown symbols.

\end{promptbox}

\begin{promptbox}[Instruction Generation Prompt: \texttt{CosyN-400K}]{orange}
\textbf{Model: GPT-4o}\\ \\
Write the general TASK to write a code for plotting the given data. \\ \\
The top two rows of the data are included in the code comments, showing the CSV structure, and the result of the generated plot image is given at the end. \\ \\
Split task into four parts:\\
1. Setup (describe programming language and libraries required to generate the plot).\\
2. Data Description (some short description of the given data).\\
3. Plot Description (describe the structural layout of the plot, without referencing libraries or function names. Begin with “Generate...” or “Create...”).\\
4. Plot Style Description (describe the visual styling aspects of the plot, without referencing libraries or function).\\ \\
CODE:
\texttt{\{code\}} \\ \\
Each part of the task must start on a new line, numbered 1 through 4. Use plain text only. Do not include any markdown symbols.

\end{promptbox}

%% file: appendix/self_debug_results.tex
\newpage
\section{Breakdown Results in Self-Debug Mode Evaluation}
\label{sec: breakdown_self_debug_results}
In this section, we provide a breakdown of model performance under the self-debug setting. For each visualization library, we report execution pass rates across up to three rounds of automatic correction, grouped by model series. 

\subsection{Matplotlib}

\begin{table}[ht]
\centering
\resizebox{0.6\linewidth}{!}{
\begin{tabular}{lcccc}
\toprule
\multirow{2}{*}{\textbf{Model}} & \multirow{2}{*}{\textbf{Normal}} & \multicolumn{3}{c}{\textbf{Self-Debug Attempt}} \\
& & \textbf{Round 1} & \textbf{Round 2} & \textbf{Round 3} \\ \midrule
GPT-4o & 94.9 & 97.7 & 99.4 & 99.4 \\
GPT-4o-mini & 88.6 & 96.6 & 97.7 & 97.7 \\
\noalign{\vskip 2pt}
\hdashline
\noalign{\vskip 2pt}
Llama-3.2-3B-Instruct & 65.1 & 76.6 & 80.0 & 81.7 \\
Qwen2.5-3B-Instruct & 74.3 & 79.4 & 82.9 & 84.6 \\
Qwen2.5-Coder-3B-Instruct & 71.4 & 74.9 & 76.6 & 76.6 \\
VisCoder-3B & 81.7 & 83.4 & 85.1 & 85.1 \\
\noalign{\vskip 2pt}
\hdashline
\noalign{\vskip 2pt}
Llama-3.1-8B-Instruct & 81.1 & 89.7 & 92.6 & 93.7 \\
Qwen2.5-7B-Instruct & 77.1 & 83.4 & 88.0 & 89.7 \\
Qwen2.5-Coder-7B-Instruct & 78.3 & 82.9 & 83.4 & 83.4 \\
VisCoder-7B & 87.4 & 90.9 & 91.4 & 91.4 \\
\bottomrule
\end{tabular}
}
\caption{\centering Execution pass rates (\%) on Matplotlib tasks under the normal and self-debug settings. Models that fail initially are allowed up to three rounds of automatic correction. \newline  \hyperref[list:list_of_appendix]{[Back to Appendix Contents]}}
\label{tab: breakdown_self_debug_matplotlib}
\end{table}

\subsection{Seaborn}

\begin{table}[ht]
\centering
\resizebox{0.6\linewidth}{!}{
\begin{tabular}{lcccc}
\toprule
\multirow{2}{*}{\textbf{Model}} & \multirow{2}{*}{\textbf{Normal}} & \multicolumn{3}{c}{\textbf{Self-Debug Attempt}} \\
& & \textbf{Round 1} & \textbf{Round 2} & \textbf{Round 3} \\ \midrule
GPT-4o & 83.4 & 90.3 & 92.6 & 92.6 \\
GPT-4o-mini & 62.3 & 69.1 & 70.9 & 72.0 \\
\noalign{\vskip 2pt}
\hdashline
\noalign{\vskip 2pt}
Llama-3.2-3B-Instruct & 30.9 & 64.6 & 72.0 & 74.9 \\
Qwen2.5-3B-Instruct & 58.3 & 64.0 & 73.7 & 75.4 \\
Qwen2.5-Coder-3B-Instruct & 58.3 & 65.7 & 68.0 & 68.0 \\
VisCoder-3B & 73.7 & 77.7 & 78.3 & 78.3 \\
\noalign{\vskip 2pt}
\hdashline
\noalign{\vskip 2pt}
Llama-3.1-8B-Instruct & 65.7 & 78.9 & 84.6 & 90.3 \\
Qwen2.5-7B-Instruct & 66.3 & 79.4 & 85.7 & 89.7 \\
Qwen2.5-Coder-7B-Instruct & 68.6 & 82.3 & 84.6 & 86.3 \\
VisCoder-7B & 76.6 & 86.9 & 89.7 & 90.3 \\
\bottomrule
\end{tabular}
}
\caption{\centering Execution pass rates (\%) on Seaborn tasks under the normal and self-debug settings. All models undergo up to three rounds of automatic correction after initial failure.\newline \hyperref[list:list_of_appendix]{[Back to Appendix Contents]}}
\label{tab: breakdown_self_debug_seaborn}
\end{table}

\newpage
\clearpage
\subsection{Plotly}

\begin{table}[ht]
\centering
\resizebox{0.6\linewidth}{!}{
\begin{tabular}{lcccc}
\toprule
\multirow{2}{*}{\textbf{Model}} & \multirow{2}{*}{\textbf{Normal}} & \multicolumn{3}{c}{\textbf{Self-Debug Attempt}} \\
& & \textbf{Round 1} & \textbf{Round 2} & \textbf{Round 3} \\ \midrule
GPT-4o & 77.7 & 92.0 & 95.4 & 97.7 \\
GPT-4o-mini & 69.1 & 88.0 & 96.0 & 97.7 \\
\noalign{\vskip 2pt}
\hdashline
\noalign{\vskip 2pt}
Llama-3.2-3B-Instruct & 13.1 & 20.6 & 24.0 & 28.0 \\
Qwen2.5-3B-Instruct & 30.9 & 36.0 & 42.3 & 48.0 \\
Qwen2.5-Coder-3B-Instruct & 27.4 & 34.9 & 36.0 & 36.0 \\
VisCoder-3B & 60.6 & 64.6 & 64.6 & 64.6 \\
\noalign{\vskip 2pt}
\hdashline
\noalign{\vskip 2pt}
Llama-3.1-8B-Instruct & 30.9 & 43.4 & 53.7 & 58.3 \\
Qwen2.5-7B-Instruct & 56.0 & 66.3 & 72.6 & 77.1 \\
Qwen2.5-Coder-7B-Instruct & 48.0 & 57.7 & 68.6 & 71.4 \\
VisCoder-7B & 74.3 & 80.0 & 81.7 & 81.7 \\
\bottomrule
\end{tabular}
}
\caption{\centering Execution pass rates (\%) on Plotly tasks under the normal and self-debug settings. All models undergo up to three rounds of automatic correction after initial failure.\newline \hyperref[list:list_of_appendix]{[Back to Appendix Contents]}}
\label{tab: breakdown_self_debug_plotly}
\end{table}

%% file: appendix/error_analysis_results.tex
\newpage
\clearpage
\section{Breakdown Results by Error Type}
\label{sec: breakdown_self_error_analysis}
In this section, we provide a detailed breakdown of execution error types across model families, plotting libraries, and self-debugging rounds. For each model series, we report the number of Python exceptions observed under default execution and across up to three rounds of automatic correction.

\subsection{VisCoder Series}

\begin{table}[ht]
\centering
\resizebox{\linewidth}{!}{
\begin{tabular}{lcccclcccclcccc}
\toprule
\multirow{2}{*}{\textbf{Error Type}} & \multicolumn{5}{c}{\textbf{Matplotlib}} & \multicolumn{5}{c}{\textbf{Seaborn}} & \multicolumn{4}{c}{\textbf{Plotly}} \\
 & \textbf{Normal} & \textbf{Round 1} & \textbf{Round 2} & \textbf{Round 3} &  & \textbf{Normal} & \textbf{Round 1} & \textbf{Round 2} & \textbf{Round 3} &  & \textbf{Normal} & \textbf{Round 1} & \textbf{Round 2} & \textbf{Round 3} \\ \midrule
AttributeError & 5 & 2 & 2 & 2 &  & 15 & 3 & 2 & 2 &  & 5 & 1 & 1 & 1 \\
AxisError & - & - & - & - &  & 1 & 1 & 1 & 1 &  & - & - & - & - \\
ImportError & 1 & 0 & 0 & 0 &  & 1 & 0 & 0 & 0 &  & - & - & - & - \\
IndexError & 1 & 0 & 0 & 0 &  & - & - & - & - &  & 1 & 1 & 1 & 1 \\
KeyError & 1 & 2 & 1 & 1 &  & - & - & - & - &  & 0 & 1 & 1 & 2 \\
KeyboardInterrupt & 1 & 1 & 1 & 1 &  & 2 & 1 & 1 & 1 &  & 1 & 1 & 1 & 1 \\
NameError & - & - & - & - &  & 5 & 4 & 2 & 1 &  & - & - & - & - \\
OSError & 1 & 1 & 1 & 1 &  & 1 & 1 & 1 & 1 &  & 1 & 1 & 1 & 1 \\
SyntaxError & 1 & 0 & 0 & 0 &  & - & - & - & - &  & 5 & 3 & 2 & 2 \\
TypeError & 7 & 5 & 5 & 5 &  & 8 & 5 & 4 & 4 &  & 3 & 1 & 1 & 1 \\
ValueError & 4 & 5 & 5 & 5 &  & 8 & 8 & 7 & 7 &  & 29 & 26 & 24 & 23 \\
\noalign{\vskip 4pt}
\hdashline
\noalign{\vskip 4pt}
\textbf{Total Errors} & 22 & 16 & 15 & 15 &  & 41 & 23 & 18 & 17 &  & 45 & 35 & 32 & 32 \\
\bottomrule
\end{tabular}
}
\caption{\centering Distribution of execution errors for VisCoder-7B across Matplotlib, Seaborn, and Plotly. Each column shows error counts at different self-debugging rounds after initial failure.\newline \hyperref[list:list_of_appendix]{[Back to Appendix Contents]}}
\label{tab:error_analysis_vis_coder_7b}
\end{table}

\begin{table}[ht]
\centering
\resizebox{\linewidth}{!}{
\begin{tabular}{lcccclcccclcccc}
\toprule
\multirow{2}{*}{\textbf{Error Type}} & \multicolumn{5}{c}{\textbf{Matplotlib}} & \multicolumn{5}{c}{\textbf{Seaborn}} & \multicolumn{4}{c}{\textbf{Plotly}} \\
 & \textbf{Normal} & \textbf{Round 1} & \textbf{Round 2} & \textbf{Round 3} &  & \textbf{Normal} & \textbf{Round 1} & \textbf{Round 2} & \textbf{Round 3} &  & \textbf{Normal} & \textbf{Round 1} & \textbf{Round 2} & \textbf{Round 3} \\ \midrule
AttributeError & 7 & 3 & 2 & 2 &  & 20 & 10 & 10 & 9 &  & 4 & 4 & 3 & 2 \\
ImportError & - & - & - & - &  & - & - & - & - &  & 1 & 1 & 1 & 0 \\
IndexError & 2 & 2 & 2 & 2 &  & 4 & 3 & 3 & 3 &  & 1 & 1 & 1 & 1 \\
KeyError & 2 & 3 & 3 & 3 &  & 3 & 4 & 4 & 4 &  & - & - & - & - \\
KeyboardInterrupt & 0 & 0 & 1 & 4 &  & 2 & 2 & 2 & 3 &  & 5 & 4 & 4 & 29 \\
NameError & 0 & 1 & 0 & 0 &  & 1 & 1 & 1 & 2 &  & 0 & 2 & 0 & 0 \\
OSError & - & - & - & - &  & 1 & 1 & 1 & 1 &  & - & - & - & - \\
SyntaxError & 3 & 3 & 2 & 0 &  & 1 & 1 & 1 & 0 &  & 8 & 6 & 6 & 1 \\
TypeError & 5 & 5 & 4 & 4 &  & 2 & 4 & 3 & 3 &  & 9 & 6 & 6 & 6 \\
ValueError & 13 & 12 & 12 & 11 &  & 12 & 13 & 13 & 13 &  & 41 & 38 & 41 & 23 \\
\noalign{\vskip 4pt}
\hdashline
\noalign{\vskip 4pt}
\textbf{Total Errors} & 32 & 29 & 26 & 26 &  & 46 & 39 & 38 & 38 &  & 69 & 62 & 62 & 62 \\
\bottomrule
\end{tabular}
}
\caption{\centering Distribution of execution errors for VisCoder-3B across Matplotlib, Seaborn, and Plotly. Each column shows error counts at different self-debugging rounds after initial failure.\newline \hyperref[list:list_of_appendix]{[Back to Appendix Contents]}}
\end{table}

\newpage
\clearpage
\subsection{GPT Series}

\begin{table}[ht]
\centering
\resizebox{\linewidth}{!}{
\begin{tabular}{lcccclcccclcccc}
\toprule
\multirow{2}{*}{\textbf{Error Type}} & \multicolumn{5}{c}{\textbf{Matplotlib}} & \multicolumn{5}{c}{\textbf{Seaborn}} & \multicolumn{4}{c}{\textbf{Plotly}} \\
 & \textbf{Normal} & \textbf{Round 1} & \textbf{Round 2} & \textbf{Round 3} &  & \textbf{Normal} & \textbf{Round 1} & \textbf{Round 2} & \textbf{Round 3} &  & \textbf{Normal} & \textbf{Round 1} & \textbf{Round 2} & \textbf{Round 3} \\ \midrule
AttributeError & - & - & - & - &  & - & - & - & - &  & 4 & 1 & 0 & 0 \\
Exception & - & - & - & - &  & 1 & 0 & 0 & 0 &  & 1 & 0 & 0 & 0 \\
IndexError & 1 & 0 & 0 & 0 &  & 1 & 1 & 0 & 0 &  & - & - & - & - \\
KeyError & - & - & - & - &  & - & - & - & - &  & 1 & 1 & 0 & 0 \\
KeyboardInterrupt & - & - & - & - &  & - & - & - & - &  & 2 & 1 & 2 & 2 \\
ModuleNotFoundError & 1 & 0 & 0 & 0 &  & 1 & 0 & 0 & 0 &  & - & - & - & - \\
NameError & - & - & - & - &  & 14 & 12 & 13 & 13 &  & 2 & 0 & 0 & 0 \\
RuntimeError & - & - & - & - &  & 1 & 0 & 0 & 0 &  & - & - & - & - \\
SyntaxError & - & - & - & - &  & - & - & - & - &  & 2 & 0 & 0 & 0 \\
TypeError & 2 & 1 & 0 & 0 &  & 6 & 1 & 0 & 0 &  & 3 & 1 & 1 & 0 \\
ValueError & 5 & 3 & 1 & 1 &  & 5 & 3 & 0 & 0 &  & 24 & 10 & 5 & 2 \\
\noalign{\vskip 4pt}
\hdashline
\noalign{\vskip 4pt}
\textbf{Total Errors} & 9 & 4 & 1 & 1 &  & 29 & 17 & 13 & 13 &  & 39 & 14 & 8 & 4 \\
\bottomrule
\end{tabular}
}
\caption{\centering Distribution of execution errors for GPT-4o across Matplotlib, Seaborn, and Plotly. Each column shows error counts at different self-debugging rounds after initial failure.\newline \hyperref[list:list_of_appendix]{[Back to Appendix Contents]}}
\end{table}

\begin{table}[ht]
\centering
\resizebox{\linewidth}{!}{
\begin{tabular}{lcccclcccclcccc}
\toprule
\multirow{2}{*}{\textbf{Error Type}} & \multicolumn{5}{c}{\textbf{Matplotlib}} & \multicolumn{5}{c}{\textbf{Seaborn}} & \multicolumn{4}{c}{\textbf{Plotly}} \\
 & \textbf{Normal} & \textbf{Round 1} & \textbf{Round 2} & \textbf{Round 3} &  & \textbf{Normal} & \textbf{Round 1} & \textbf{Round 2} & \textbf{Round 3} &  & \textbf{Normal} & \textbf{Round 1} & \textbf{Round 2} & \textbf{Round 3} \\ \midrule
AttributeError & 3 & 0 & 0 & 0 &  & 2 & 0 & 0 & 0 &  & 13 & 2 & 0 & 0 \\
Exception & 1 & 0 & 0 & 0 &  & 2 & 0 & 1 & 0 &  & 1 & 1 & 1 & 1 \\
FileNotFoundError & 1 & 0 & 0 & 0 &  & 1 & 1 & 0 & 0 &  & - & - & - & - \\
ImportError & 1 & 0 & 0 & 0 &  & - & - & - & - &  & - & - & - & - \\
IndexError & 0 & 1 & 1 & 1 &  & 1 & 1 & 1 & 1 &  & 0 & 2 & 1 & 1 \\
KeyError & 2 & 1 & 0 & 0 &  & 1 & 1 & 0 & 0 &  & - & - & - & - \\
KeyboardInterrupt & - & - & - & - &  & 1 & 0 & 0 & 0 &  & 1 & 0 & 0 & 0 \\
ModuleNotFoundError & 1 & 0 & 0 & 0 &  & - & - & - & - &  & 2 & 0 & 0 & 0 \\
NameError & 1 & 0 & 0 & 0 &  & 43 & 48 & 47 & 47 &  & 11 & 0 & 0 & 0 \\
TypeError & 1 & 1 & 1 & 0 &  & 4 & 1 & 0 & 0 &  & 5 & 1 & 1 & 0 \\
ValueError & 9 & 3 & 2 & 3 &  & 11 & 2 & 2 & 1 &  & 21 & 15 & 4 & 2 \\
\noalign{\vskip 4pt}
\hdashline
\noalign{\vskip 4pt}
\textbf{Total Errors} & 20 & 6 & 4 & 4 &  & 66 & 54 & 51 & 49 &  & 54 & 21 & 7 & 4 \\
\bottomrule
\end{tabular}
}
\caption{\centering Distribution of execution errors for GPT-4o-mini across Matplotlib, Seaborn, and Plotly. Each column shows error counts at different self-debugging rounds after initial failure.\newline \hyperref[list:list_of_appendix]{[Back to Appendix Contents]}}
\end{table}

\newpage
\clearpage
\subsection{Qwen2.5 Series}
 
\begin{table}[ht]
\centering
\resizebox{\linewidth}{!}{
\begin{tabular}{lcccclcccclcccc}
\toprule
\multirow{2}{*}{\textbf{Error Type}} & \multicolumn{5}{c}{\textbf{Matplotlib}} & \multicolumn{5}{c}{\textbf{Seaborn}} & \multicolumn{4}{c}{\textbf{Plotly}} \\
 & \textbf{Normal} & \textbf{Round 1} & \textbf{Round 2} & \textbf{Round 3} &  & \textbf{Normal} & \textbf{Round 1} & \textbf{Round 2} & \textbf{Round 3} &  & \textbf{Normal} & \textbf{Round 1} & \textbf{Round 2} & \textbf{Round 3} \\ \midrule
AttributeError & 12 & 9 & 9 & 9 &  & 17 & 8 & 7 & 7 &  & 8 & 5 & 3 & 3 \\
FileNotFoundError & - & - & - & - &  & 0 & 1 & 1 & 1 &  & - & - & - & - \\
ImportError & 1 & 0 & 0 & 0 &  & - & - & - & - &  & - & - & - & - \\
IndexError & 1 & 1 & 1 & 1 &  & 1 & 0 & 0 & 0 &  & 1 & 1 & 1 & 1 \\
KeyError & 3 & 4 & 3 & 3 &  & 2 & 1 & 1 & 1 &  & 14 & 17 & 0 & 0 \\
KeyboardInterrupt & 1 & 1 & 1 & 1 &  & 1 & 2 & 2 & 2 &  & 2 & 1 & 1 & 1 \\
ModuleNotFoundError & - & - & - & - &  & - & - & - & - &  & 0 & 1 & 1 & 1 \\
NameError & 1 & 0 & 0 & 0 &  & 17 & 8 & 5 & 3 &  & - & - & - & - \\
SyntaxError & - & - & - & - &  & - & - & - & - &  & 6 & 4 & 7 & 5 \\
TypeError & 9 & 7 & 7 & 7 &  & 8 & 5 & 4 & 4 &  & 7 & 1 & 1 & 1 \\
ValueError & 10 & 8 & 8 & 8 &  & 9 & 6 & 7 & 6 &  & 53 & 44 & 41 & 38 \\
\noalign{\vskip 4pt}
\hdashline
\noalign{\vskip 4pt}
\textbf{Total Errors} & 38 & 30 & 29 & 29 &  & 55 & 31 & 27 & 24 &  & 91 & 74 & 55 & 50 \\
\bottomrule
\end{tabular}
}
\caption{\centering Distribution of execution errors for Qwen2.5-Coder-7B-Instruct across Matplotlib, Seaborn, and Plotly. Each column shows error counts at different self-debugging rounds after initial failure.\newline \hyperref[list:list_of_appendix]{[Back to Appendix Contents]}}
\end{table}

\begin{table}[ht]
\centering
\resizebox{\linewidth}{!}{
\begin{tabular}{lcccclcccclcccc}
\toprule
\multirow{2}{*}{\textbf{Error Type}} & \multicolumn{5}{c}{\textbf{Matplotlib}} & \multicolumn{5}{c}{\textbf{Seaborn}} & \multicolumn{4}{c}{\textbf{Plotly}} \\
 & \textbf{Normal} & \textbf{Round 1} & \textbf{Round 2} & \textbf{Round 3} &  & \textbf{Normal} & \textbf{Round 1} & \textbf{Round 2} & \textbf{Round 3} &  & \textbf{Normal} & \textbf{Round 1} & \textbf{Round 2} & \textbf{Round 3} \\ \midrule
AttributeError & 10 & 7 & 7 & 3 &  & 14 & 4 & 5 & 2 &  & 9 & 7 & 4 & 3 \\
FileNotFoundError & 1 & 0 & 0 & 0 &  & - & - & - & - &  & - & - & - & - \\
ImportError & - & - & - & - &  & 0 & 1 & 0 & 0 &  & - & - & - & - \\
IndexError & - & - & - & - &  & 2 & 3 & 1 & 1 &  & - & - & - & - \\
KeyError & 2 & 1 & 1 & 1 &  & 3 & 0 & 0 & 0 &  & - & - & - & - \\
KeyboardInterrupt & - & - & - & - &  & 0 & 1 & 0 & 0 &  & 1 & 1 & 2 & 2 \\
ModuleNotFoundError & - & - & - & - &  & 1 & 1 & 0 & 0 &  & - & - & - & - \\
NameError & 1 & 1 & 0 & 0 &  & 13 & 8 & 6 & 4 &  & 10 & 11 & 9 & 10 \\
RecursionError & 1 & 1 & 1 & 1 &  & - & - & - & - &  & - & - & - & - \\
OSError & - & - & - & - &  & 1 & 1 & 1 & 1 &  & - & - & - & - \\
RuntimeError & - & - & - & - &  & 1 & 1 & 1 & 0 &  & - & - & - & - \\
SyntaxError & 1 & 0 & 0 & 0 &  & 1 & 1 & 0 & 0 &  & 4 & 1 & 2 & 1 \\
TypeError & 12 & 5 & 3 & 4 &  & 8 & 4 & 2 & 1 &  & 2 & 1 & 2 & 1 \\
ValueError & 12 & 14 & 9 & 9 &  & 15 & 11 & 9 & 9 &  & 51 & 38 & 29 & 23 \\
\noalign{\vskip 4pt}
\hdashline
\noalign{\vskip 4pt}
\textbf{Total Errors} & 40 & 29 & 21 & 18 &  & 59 & 36 & 25 & 18 &  & 77 & 59 & 48 & 40 \\
\bottomrule
\end{tabular}
}
\caption{\centering Distribution of execution errors for Qwen2.5-7B-Instruct across Matplotlib, Seaborn, and Plotly. Each column shows error counts at different self-debugging rounds after initial failure.\newline \hyperref[list:list_of_appendix]{[Back to Appendix Contents]}}
\end{table}

\newpage
\clearpage

\begin{table}[ht]
\centering
\resizebox{\linewidth}{!}{
\begin{tabular}{lcccclcccclcccc}
\toprule
\multirow{2}{*}{\textbf{Error Type}} & \multicolumn{5}{c}{\textbf{Matplotlib}} & \multicolumn{5}{c}{\textbf{Seaborn}} & \multicolumn{4}{c}{\textbf{Plotly}} \\
 & \textbf{Normal} & \textbf{Round 1} & \textbf{Round 2} & \textbf{Round 3} &  & \textbf{Normal} & \textbf{Round 1} & \textbf{Round 2} & \textbf{Round 3} &  & \textbf{Normal} & \textbf{Round 1} & \textbf{Round 2} & \textbf{Round 3} \\ \midrule
AssertionError & 1 & 1 & 1 & 1 &  & - & - & - & - &  & - & - & - & - \\
AttributeError & 14 & 9 & 8 & 8 &  & 32 & 23 & 21 & 21 &  & 31 & 13 & 12 & 11 \\
FileNotFoundError & 2 & 1 & 1 & 1 &  & - & - & - & - &  & - & - & - & - \\
IndexError & 3 & 3 & 3 & 3 &  & 4 & 4 & 4 & 4 &  & - & - & - & - \\
KeyError & 1 & 1 & 1 & 1 &  & 3 & 1 & 1 & 1 &  & 1 & 2 & 1 & 1 \\
KeyboardInterrupt & 1 & 1 & 1 & 3 &  & 1 & 0 & 0 & 1 &  & 2 & 2 & 5 & 36 \\
NameError & - & - & - & - &  & 6 & 1 & 0 & 0 &  & 1 & 2 & 1 & 2 \\
SyntaxError & 3 & 4 & 3 & 1 &  & 0 & 1 & 1 & 0 &  & 13 & 16 & 13 & 3 \\
TypeError & 11 & 10 & 10 & 10 &  & 7 & 9 & 9 & 9 &  & 37 & 23 & 25 & 24 \\
ValueError & 14 & 14 & 13 & 13 &  & 20 & 21 & 20 & 20 &  & 42 & 56 & 55 & 35 \\
\noalign{\vskip 4pt}
\hdashline
\noalign{\vskip 4pt}
\textbf{Total Errors} & 50 & 44 & 41 & 41 &  & 73 & 60 & 56 & 56 &  & 127 & 114 & 112 & 112 \\
\bottomrule
\end{tabular}
}
\caption{\centering Distribution of execution errors for Qwen2.5-Coder-3B-Instruct across Matplotlib, Seaborn, and Plotly. Each column shows error counts at different self-debugging rounds after initial failure.\newline \hyperref[list:list_of_appendix]{[Back to Appendix Contents]}}
\end{table}

\begin{table}[ht]
\centering
\resizebox{\linewidth}{!}{
\begin{tabular}{lcccclcccclcccc}
\toprule
\multirow{2}{*}{\textbf{Error Type}} & \multicolumn{5}{c}{\textbf{Matplotlib}} & \multicolumn{5}{c}{\textbf{Seaborn}} & \multicolumn{4}{c}{\textbf{Plotly}} \\
 & \textbf{Normal} & \textbf{Round 1} & \textbf{Round 2} & \textbf{Round 3} &  & \textbf{Normal} & \textbf{Round 1} & \textbf{Round 2} & \textbf{Round 3} &  & \textbf{Normal} & \textbf{Round 1} & \textbf{Round 2} & \textbf{Round 3} \\ \midrule
AttributeError & 11 & 9 & 4 & 4 &  & 29 & 19 & 13 & 11 &  & 32 & 20 & 13 & 8 \\
FileNotFoundError & 1 & 0 & 0 & 0 &  & 6 & 6 & 5 & 5 &  & - & - & - & - \\
ImportError & 1 & 1 & 1 & 1 &  & - & - & - & - &  & - & - & - & - \\
IndexError & - & - & - & - &  & 1 & 1 & 0 & 1 &  & 0 & 0 & 0 & 1 \\
KeyError & 4 & 3 & 2 & 1 &  & 4 & 2 & 1 & 1 &  & 2 & 2 & 2 & 2 \\
KeyboardInterrupt & 2 & 2 & 2 & 3 &  & 2 & 1 & 2 & 2 &  & 1 & 1 & 2 & 36 \\
NameError & 2 & 1 & 1 & 1 &  & 2 & 0 & 0 & 1 &  & 1 & 1 & 3 & 2 \\
NotImplementedError & 1 & 0 & 1 & 0 &  & - & - & - & - &  & - & - & - & - \\
RuntimeError & 1 & 1 & 1 & 1 &  & 1 & 1 & 1 & 1 &  & - & - & - & - \\
SyntaxError & 2 & 1 & 1 & 2 &  & 3 & 3 & 0 & 0 &  & 14 & 15 & 13 & 3 \\
TypeError & 11 & 8 & 5 & 4 &  & 10 & 11 & 6 & 4 &  & 18 & 15 & 11 & 5 \\
ValueError & 9 & 10 & 12 & 10 &  & 15 & 19 & 18 & 17 &  & 53 & 58 & 57 & 34 \\
\noalign{\vskip 4pt}
\hdashline
\noalign{\vskip 4pt}
\textbf{Total Errors} & 45 & 36 & 30 & 27 &  & 73 & 63 & 46 & 43 &  & 121 & 112 & 101 & 91 \\
\bottomrule
\end{tabular}
}
\caption{\centering Distribution of execution errors for Qwen2.5-3B-Instruct across Matplotlib, Seaborn, and Plotly. Each column shows error counts at different self-debugging rounds after initial failure.\newline \hyperref[list:list_of_appendix]{[Back to Appendix Contents]}}
\end{table}

\newpage
\clearpage
\subsection{LLaMA Series}

\begin{table}[ht]
\centering
\resizebox{\linewidth}{!}{
\begin{tabular}{lcccclcccclcccc}
\toprule
\multirow{2}{*}{\textbf{Error Type}} & \multicolumn{5}{c}{\textbf{Matplotlib}} & \multicolumn{5}{c}{\textbf{Seaborn}} & \multicolumn{4}{c}{\textbf{Plotly}} \\
 & \textbf{Normal} & \textbf{Round 1} & \textbf{Round 2} & \textbf{Round 3} &  & \textbf{Normal} & \textbf{Round 1} & \textbf{Round 2} & \textbf{Round 3} &  & \textbf{Normal} & \textbf{Round 1} & \textbf{Round 2} & \textbf{Round 3} \\ \midrule
AttributeError & 10 & 3 & 1 & 1 &  & 21 & 6 & 8 & 3 &  & 27 & 20 & 10 & 9 \\
FileNotFoundError & 1 & 0 & 0 & 0 &  & 2 & 3 & 0 & 0 &  & - & - & - & - \\
IndexError & 0 & 1 & 1 & 0 &  & 2 & 1 & 0 & 0 &  & 2 & 0 & 3 & 2 \\
KeyError & 1 & 1 & 2 & 2 &  & 1 & 1 & 2 & 1 &  & 2 & 2 & 3 & 2 \\
KeyboardInterrupt & - & - & - & - &  & 0 & 4 & 1 & 2 &  & - & - & - & - \\
NameError & 1 & 0 & 1 & 0 &  & 5 & 6 & 5 & 2 &  & 1 & 2 & 1 & 1 \\
RuntimeError & - & - & - & - &  & 3 & 1 & 0 & 0 &  & - & - & - & - \\
SyntaxError & 1 & 1 & 1 & 0 &  & 0 & 1 & 1 & 0 &  & 19 & 16 & 12 & 7 \\
TypeError & 5 & 5 & 4 & 3 &  & 7 & 5 & 3 & 2 &  & 27 & 19 & 13 & 4 \\
ValueError & 14 & 7 & 3 & 4 &  & 19 & 10 & 7 & 6 &  & 43 & 40 & 38 & 38 \\
\noalign{\vskip 4pt}
\hdashline
\noalign{\vskip 4pt}
\textbf{Total Errors} & 33 & 18 & 13 & 10 &  & 60 & 38 & 27 & 16 &  & 121 & 99 & 80 & 63 \\
\bottomrule
\end{tabular}
}
\caption{\centering Distribution of execution errors for Llama-3.1-8B-Instruct across Matplotlib, Seaborn, and Plotly. Each column shows error counts at different self-debugging rounds after initial failure.\newline \hyperref[list:list_of_appendix]{[Back to Appendix Contents]}}
\end{table}

\begin{table}[ht]
\centering
\resizebox{\linewidth}{!}{
\begin{tabular}{lcccclcccclcccc}
\toprule
\multirow{2}{*}{\textbf{Error Type}} & \multicolumn{5}{c}{\textbf{Matplotlib}} & \multicolumn{5}{c}{\textbf{Seaborn}} & \multicolumn{4}{c}{\textbf{Plotly}} \\
 & \textbf{Normal} & \textbf{Round 1} & \textbf{Round 2} & \textbf{Round 3} &  & \textbf{Normal} & \textbf{Round 1} & \textbf{Round 2} & \textbf{Round 3} &  & \textbf{Normal} & \textbf{Round 1} & \textbf{Round 2} & \textbf{Round 3} \\ \midrule
AttributeError & 11 & 7 & 6 & 4 &  & 28 & 15 & 13 & 11 &  & 44 & 17 & 11 & 7 \\
FileNotFoundError & 2 & 0 & 0 & 0 &  & 4 & 1 & 0 & 0 &  & - & - & - & - \\
IndexError & 2 & 1 & 1 & 2 &  & 1 & 0 & 0 & 0 &  & 0 & 0 & 0 & 1 \\
KeyError & 1 & 4 & 1 & 1 &  & 2 & 3 & 2 & 1 &  & 1 & 1 & 1 & 1 \\
KeyboardInterrupt & 1 & 1 & 1 & 2 &  & 1 & 2 & 2 & 2 &  & 0 & 1 & 1 & 2 \\
NameError & 1 & 0 & 0 & 0 &  & 44 & 9 & 3 & 0 &  & 4 & 1 & 0 & 0 \\
SyntaxError & 3 & 2 & 1 & 0 &  & 5 & 3 & 3 & 2 &  & 21 & 24 & 22 & 24 \\
TypeError & 22 & 10 & 10 & 8 &  & 16 & 12 & 11 & 13 &  & 47 & 41 & 40 & 36 \\
UFuncTypeError & - & - & - & - &  & - & - & - & - &  & 1 & 1 & 1 & 1 \\
ValueError & 18 & 16 & 15 & 15 &  & 20 & 17 & 15 & 15 &  & 34 & 53 & 57 & 54 \\
\noalign{\vskip 4pt}
\hdashline
\noalign{\vskip 4pt}
\textbf{Total Errors} & 61 & 41 & 35 & 32 &  & 121 & 62 & 49 & 44 &  & 152 & 139 & 133 & 126 \\
\bottomrule
\end{tabular}
}
\caption{\centering Distribution of execution errors for Llama-3.2-3B-Instruct across Matplotlib, Seaborn, and Plotly. Each column shows error counts at different self-debugging rounds after initial failure.\newline \hyperref[list:list_of_appendix]{[Back to Appendix Contents]}}
\end{table}

%% file: appendix/supp_results.tex
\newpage
\clearpage

\section{Extended Results and Analyses}
\label{sec:extended_results}
In this section, we present additional results and analyses, including model scaling, baseline extensions, and dataset diversity.

\subsection{Model Scale Extension}
To address concerns regarding model scaling, we have extended our evaluation to include Qwen2.5-Coder models at 1.5B and 14B scales, in addition to the previously reported 3B and 7B settings. All models are fine-tuned on the VisCode-200K dataset using the same training configuration. Table~\ref{tab:A1} presents the results across the three plotting libraries.

\begin{table}[ht]
\centering
\resizebox{1\linewidth}{!}{
\begin{tabular}{lcccccc}
\toprule
\multirow{2}{*}{\textbf{Model}} & \multicolumn{2}{c}{\textbf{Matplotlib}} & \multicolumn{2}{c}{\textbf{Seaborn}} & \multicolumn{2}{c}{\textbf{Plotly}} \\
 & \multicolumn{1}{l}{\textbf{Exec Pass}} & \textbf{Self-Debug} & \multicolumn{1}{l}{\textbf{Exec Pass}} & \textbf{Self-Debug} & \multicolumn{1}{l}{\textbf{Exec Pass}} & \textbf{Self-Debug} \\ \midrule
Qwen-2.5-Coder-1.5B-Ins. & 66.9 & 69.1 & 56.0 & 61.1 & 26.3 & 33.1 \\
VisCoder-1.5B & \textbf{78.3} & \textbf{80.0} & \textbf{78.3} & \textbf{81.7} & \textbf{65.1} & \textbf{71.4} \\
\noalign{\vskip 2pt}
\hdashline
\noalign{\vskip 2pt}
Qwen-2.5-Coder-3B-Ins. & 71.4 & 76.6 & 58.3 & 68.0 & 27.4 & 36.0 \\
VisCoder-3B & \textbf{81.7} & \textbf{85.1} & \textbf{73.7} & \textbf{78.3} & \textbf{60.6} & \textbf{64.6} \\
\noalign{\vskip 2pt}
\hdashline
\noalign{\vskip 2pt}
Qwen-2.5-Coder-7B-Ins. & 78.3 & 83.4 & 68.6 & 86.3 & 48.0 & 71.4 \\
VisCoder-7B & \textbf{87.4} & \textbf{91.4} & \textbf{76.6} & \textbf{90.3} & \textbf{74.3} & \textbf{81.7} \\
\noalign{\vskip 2pt}
\hdashline
\noalign{\vskip 2pt}
Qwen-2.5-Coder-14B-Ins. & 82.9 & 93.7 & 76.6 & 86.9 & 56.0 & 82.9 \\
VisCoder-14B & \textbf{86.3} & \textbf{93.7} & \textbf{78.9} & \textbf{92.6} & \textbf{74.3} & \textbf{93.1} \\ \bottomrule
\end{tabular}
}
\caption{Execution pass rates (\%) with and without self-debugging across different model scales on VisCode-200K.}
\label{tab:A1}
\end{table}

These results confirm that VisCoder consistently outperforms the Qwen2.5-Coder baselines at all four scales. Performance gains are evident across all libraries, under both default and self-debug settings.

\subsection{Baseline Model Extension}
To demonstrate the generalizability of VisCode-200K, we additionally fine-tune three popular code-generation backbones: DeepSeek-Coder, CodeLLaMA, and StarCoder. Each model is trained under the same setup and evaluated on PandasPlotBench. Results are shown in Table~\ref{tab:A2}.

\begin{table}[ht]
\centering
\resizebox{1\linewidth}{!}{
\begin{tabular}{lcccccc}
\toprule
\multirow{2}{*}{\textbf{Model}} & \multicolumn{2}{c}{\textbf{Matplotlib}} & \multicolumn{2}{c}{\textbf{Seaborn}} & \multicolumn{2}{c}{\textbf{Plotly}} \\
 & \multicolumn{1}{l}{\textbf{Exec Pass}} & \textbf{Self-Debug} & \multicolumn{1}{l}{\textbf{Exec Pass}} & \textbf{Self-Debug} & \multicolumn{1}{l}{\textbf{Exec Pass}} & \textbf{Self-Debug} \\ \midrule
DeepSeek-Coder-1.3B-Ins. & 38.3 & 64.0 & 31.4 & 50.3 & 23.4 & 45.1 \\
VisCoder-DeepSeek-1.3B & \textbf{69.1} & \textbf{69.7} & \textbf{34.9} & \textbf{63.4} & \textbf{62.3} & \textbf{62.9} \\
\noalign{\vskip 2pt}
\hdashline
\noalign{\vskip 2pt}
DeepSeek-Coder-6.7B-Ins. & 80.6 & 85.7 & 41.1 & 72.6 & 48.0 & 74.9 \\
VisCoder-DeepSeek-6.7B & \textbf{88.0} & \textbf{89.1} & \textbf{68.0} & \textbf{81.1} & \textbf{65.7} & \textbf{71.4} \\
\noalign{\vskip 2pt}
\hdashline
\noalign{\vskip 2pt}
CodeLLaMA-7B-Ins. & 53.1 & 56.6 & 35.4 & 49.7 & 24.0 & 27.4 \\
VisCoder-CodeLLaMA-7B & \textbf{78.3} & \textbf{81.1} & \textbf{39.4} & \textbf{65.1} & \textbf{65.1} & \textbf{68.0} \\
\noalign{\vskip 2pt}
\hdashline
\noalign{\vskip 2pt}
CodeLLaMA-13B-Ins. & 66.9 & 69.7 & 62.3 & 72.0 & 42.9 & 49.1 \\
VisCoder-CodeLLaMA-13B & \textbf{77.1} & \textbf{81.7} & \textbf{61.7} & \textbf{71.4} & \textbf{69.1} & \textbf{71.4} \\
\noalign{\vskip 2pt}
\hdashline
\noalign{\vskip 2pt}
StarCoder2-15B-Ins. & 40.0 & 46.3 & 26.9 & 47.4 & 18.3 & 19.4 \\
VisCoder-StarCoder2-15B & \textbf{81.7} & \textbf{88.0} & \textbf{65.7} & \textbf{85.1} & \textbf{68.0} & \textbf{76.0} \\ \bottomrule
\end{tabular}
}
\caption{Execution pass rates (\%) of VisCoder fine-tuned on different code-generation backbones, compared to their respective baselines.}
\label{tab:A2}
\end{table}

These findings highlight that VisCode-200K brings consistent gains across backbones, demonstrating its broad applicability to diverse code-generation models.

\subsection{Dataset Diversity Analysis}

We analyze the library usage and visualization type distribution in VisCode-200K. The results are shown in Tables~\ref{tab:A3} and \ref{tab:A4}.

\begin{table}[ht]
\centering
\resizebox{0.41\linewidth}{!}{
\begin{tabular}{lclc}
\toprule
\textbf{Library} & \textbf{Count} & \textbf{Library} & \textbf{Count} \\ \midrule
plotly     & 62583 & altair    & 899 \\
matplotlib & 61940 & pyvista   & 271 \\
seaborn    & 30559 & holoviews & 36 \\
bokeh      & 1737  & ggplot    & 31 \\
missingno  & 25    &           &     \\ \bottomrule
\end{tabular}
}
\caption{Library usage statistics in VisCode-200K after downsampling to mitigate the dominance of matplotlib.}
\label{tab:A3}
\end{table}

To mitigate bias from the original source datasets, where matplotlib is dominant, we retain all non-matplotlib samples and randomly downsample matplotlib code. While the final distribution still reflects a high proportion of matplotlib, seaborn, and plotly, this is consistent with their prevalence in real-world usage. We have made our best effort to control distributional bias while preserving the natural frequency of commonly used libraries. All remaining samples are runtime-validated for executability and renderability.

\begin{table}[ht]
\centering
\resizebox{0.8\linewidth}{!}{
\begin{tabular}{lclclc}
\toprule
\textbf{VisType} & \textbf{Count} & \textbf{Category} & \textbf{VisType} & \textbf{Count} & \textbf{Category} \\ \midrule
line      & 95798 & Basic        & map         & 2252 & Geographic \\
scatter   & 40360 & Basic        & network     & 3194 & Graph/Network \\
bar       & 28322 & Basic        & sankey      & 321  & Graph/Network \\
area      & 2428  & Basic        & chord       & 23   & Graph/Network \\
polar     & 2408  & Basic        & dendrogram  & 72   & HierarchicalTree \\
pie       & 7889  & Basic        & treemap     & 806  & HierarchicalTree \\
grid      & 16266 & Layout/Grid  & sunburst    & 243  & HierarchicalTree \\
histogram & 8171  & Distribution & regression  & 1148 & Statistical \\
kde       & 8588  & Distribution & joint       & 304  & Statistical \\
box       & 4520  & Distribution & bubble      & 776  & Statistical \\
violin    & 1301  & Distribution & funnel      & 887  & Statistical \\
contour   & 1222  & Distribution & waterfall   & 645  & Statistical \\
hexbin    & 724   & Distribution & candlestick & 1311 & Statistical \\
count     & 2698  & Distribution & vectorfield & 729  & FlowField \\
heatmap   & 7298  & Matrix-Based & timeseries  & 878  & Temporal \\
pcolormesh& 167   & Matrix-Based & timeline    & 405  & Temporal \\
3d        & 2504  & 3D Vis      & animation   & 1100 & Temporal \\
ternary   & 698   & 3D Vis      & errorbar    & 1326 & Uncertainty/Error \\
wordcloud & 19    & Text Vis    &             &      &  \\ \bottomrule
\end{tabular}
}
\caption{Distribution of 37 visualization types in VisCode-200K, categorized into 13 groups.}
\label{tab:A4}
\end{table}

We categorize a total of 37 visualization types across 13 categories, showcasing the diversity and coverage of VisCode-200K in Python-based visualization.

%% file: appendix/case_study.tex
\newpage
\clearpage
\section{Case Study}
In this section, we present a set of representative examples from VisCoder-7B to illustrate model behavior across the three visualization libraries.

\subsection{Matplotlib: Successful Generation}
\begin{figure*}[!htbp]
    \centering
    \includegraphics[width=1\linewidth]{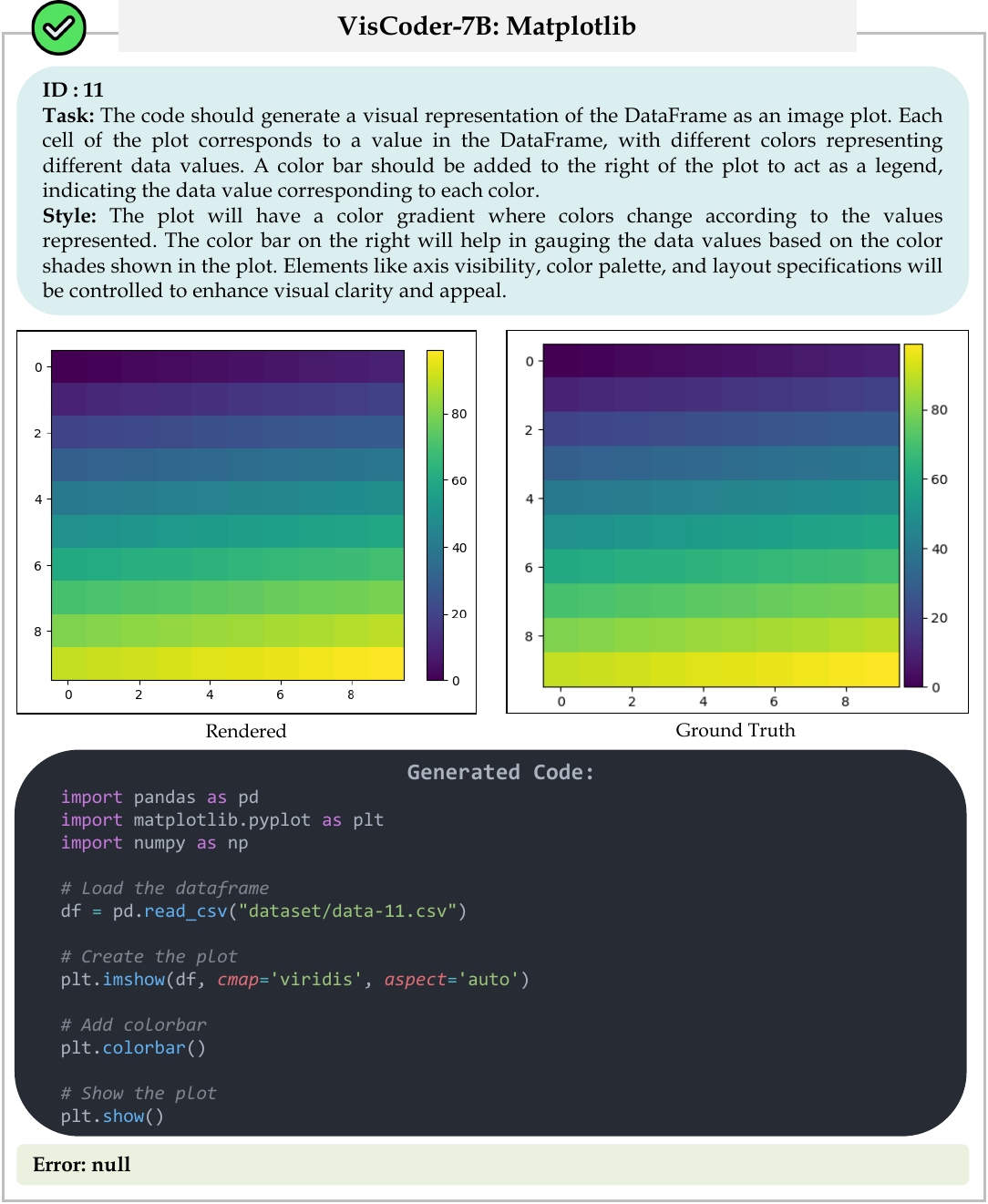}
    \caption{Example of a successful generation in \textbf{Matplotlib} (ID: 11). The model generates code that executes successfully and produces a plot consistent with the ground truth.
    \newline \centering \hyperref[list:list_of_appendix]{[Back to Appendix Contents]}}
\label{fig:case_matplotlib_correct}
\end{figure*}

\newpage
\clearpage
\subsection{Matplotlib: Self-Debug Recovery}
\begin{figure*}[!htbp]
    \centering
    \includegraphics[width=1\linewidth]{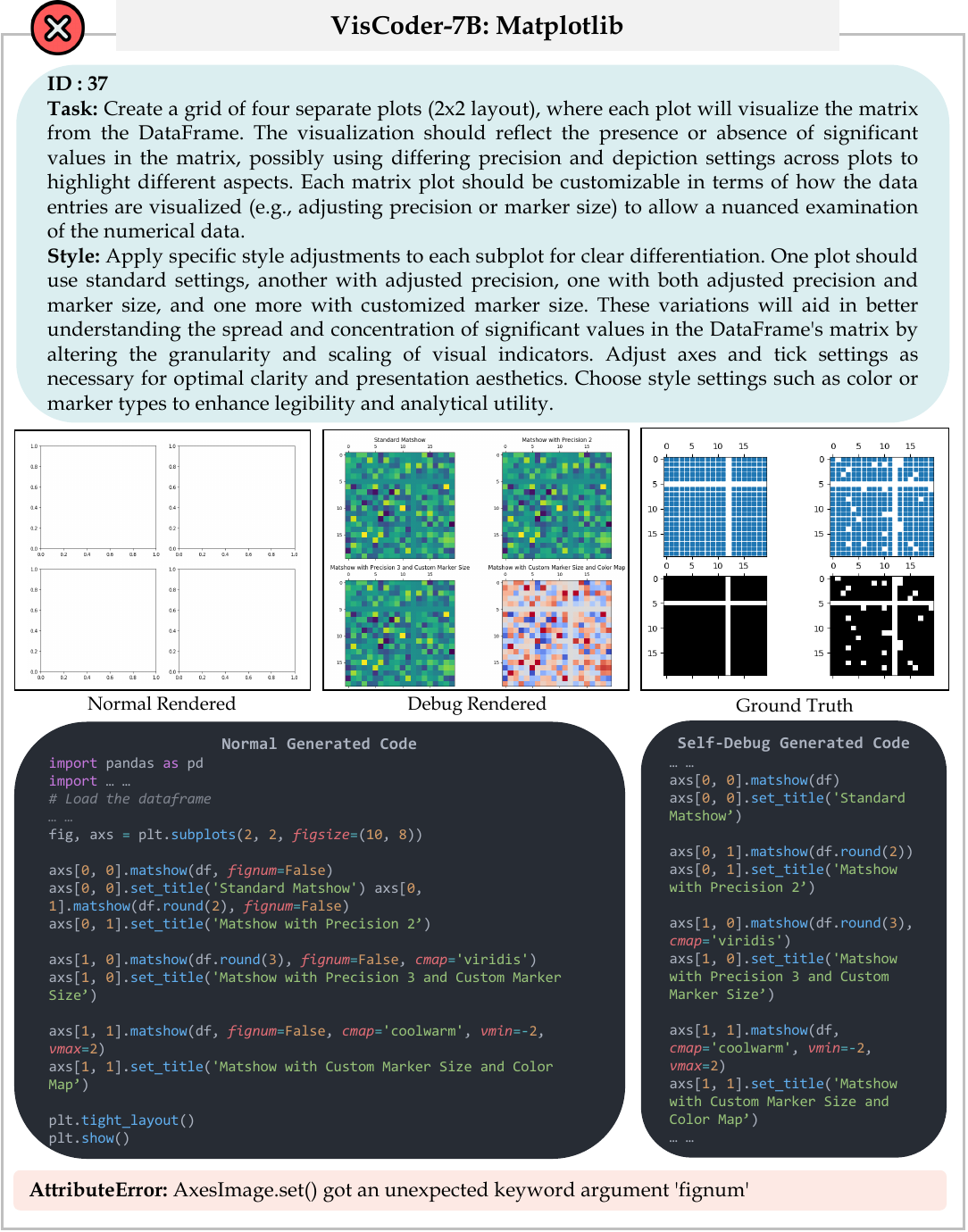}
    \caption{Example of a failed generation in \textbf{Matplotlib} (ID: 37), where the initial code raises a \textbf{AttributeError} and is resolved in the \textbf{first} round of self-debug, resulting in a corrected plot that matches the intended semantics.
    \newline \centering \hyperref[list:list_of_appendix]{[Back to Appendix Contents]}}
\label{fig:case_matplotlib_debug}
\end{figure*}

\newpage
\clearpage
\subsection{Seaborn: Successful Generation}
\begin{figure*}[!htbp]
    \centering
    \includegraphics[width=1\linewidth]{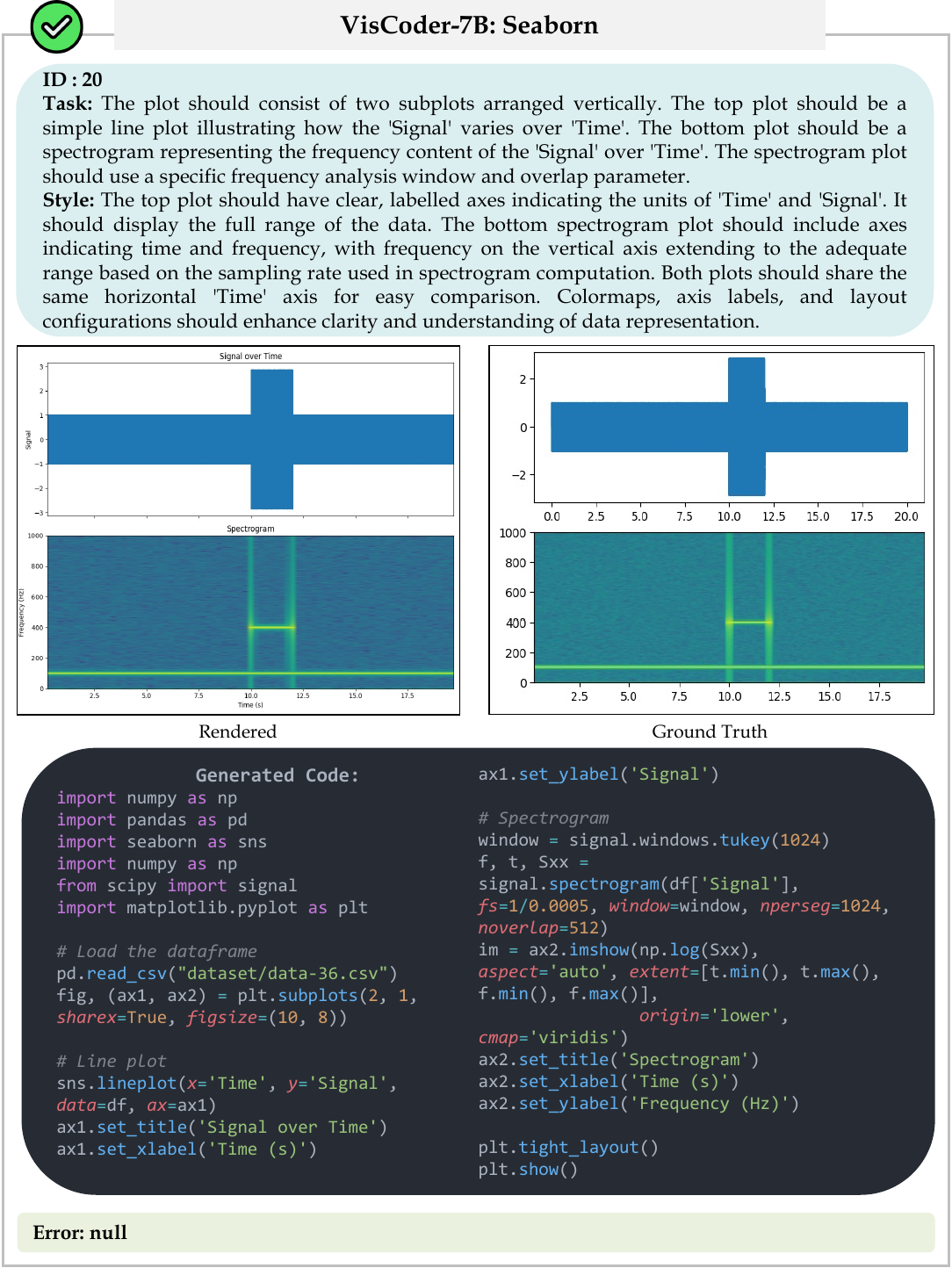}
    \caption{Example of a successful generation in \textbf{Seaborn} (ID: 20). The model generates code that executes successfully and produces a plot consistent with the ground truth.
    \newline \centering \hyperref[list:list_of_appendix]{[Back to Appendix Contents]}}
\label{fig:case_seaborn_correct}
\end{figure*}

\newpage
\clearpage
\subsection{Seaborn: Self-Debug Recovery}
\begin{figure*}[!htbp]
    \centering
    \includegraphics[width=1\linewidth]{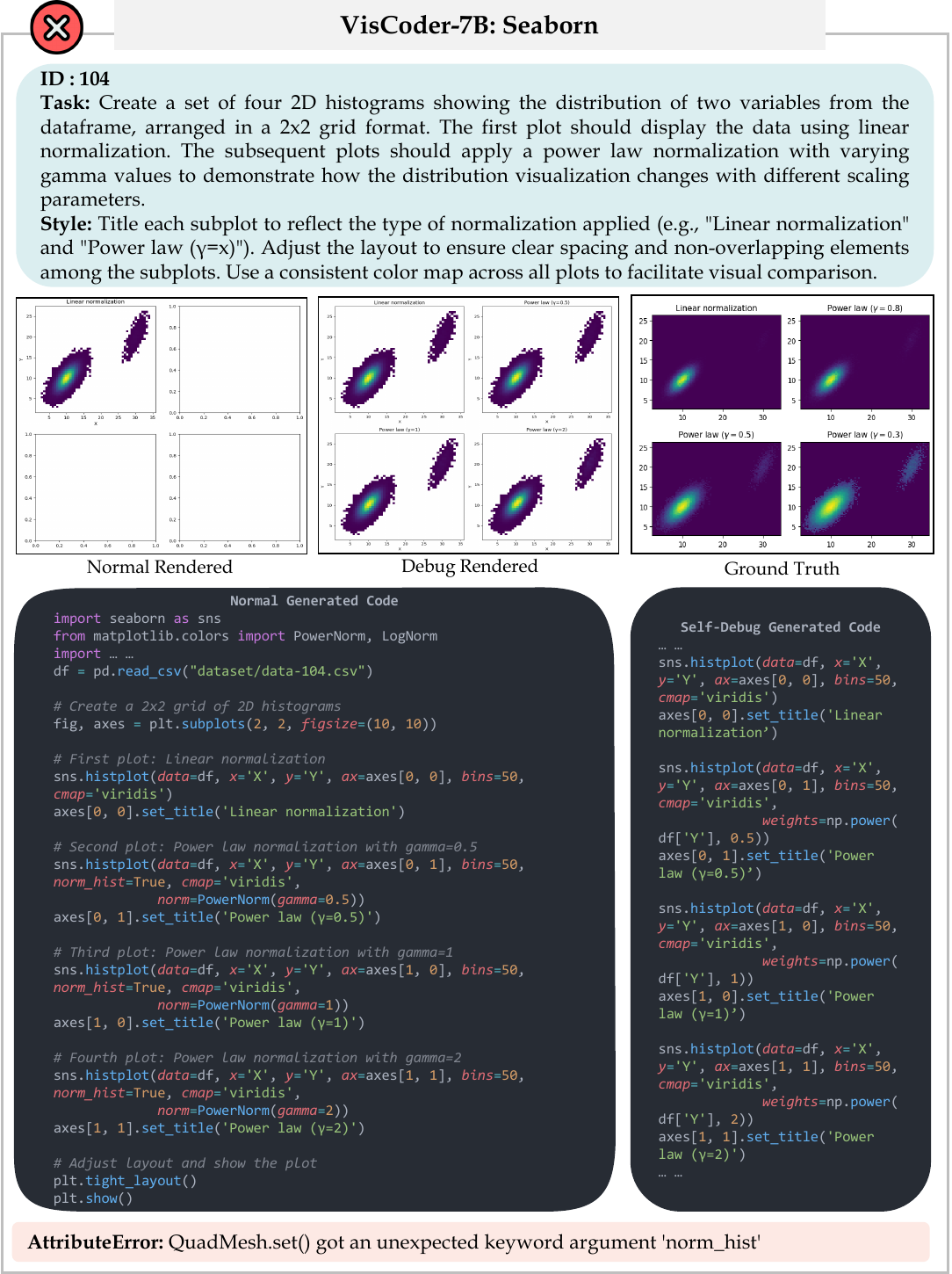}
    \caption{Example of a failed generation in \textbf{Seaborn} (ID: 104), where the initial code raises a \textbf{AttributeError} and is resolved in the \textbf{Third} round of self-debug, resulting in a corrected plot that matches the intended semantics.
    \newline \centering \hyperref[list:list_of_appendix]{[Back to Appendix Contents]}}
\label{fig:case_seaborn_debug}
\end{figure*}

\newpage
\clearpage
\subsection{Plotly: Successful Generation}
\begin{figure*}[!htbp]
    \centering
    \includegraphics[width=1\linewidth]{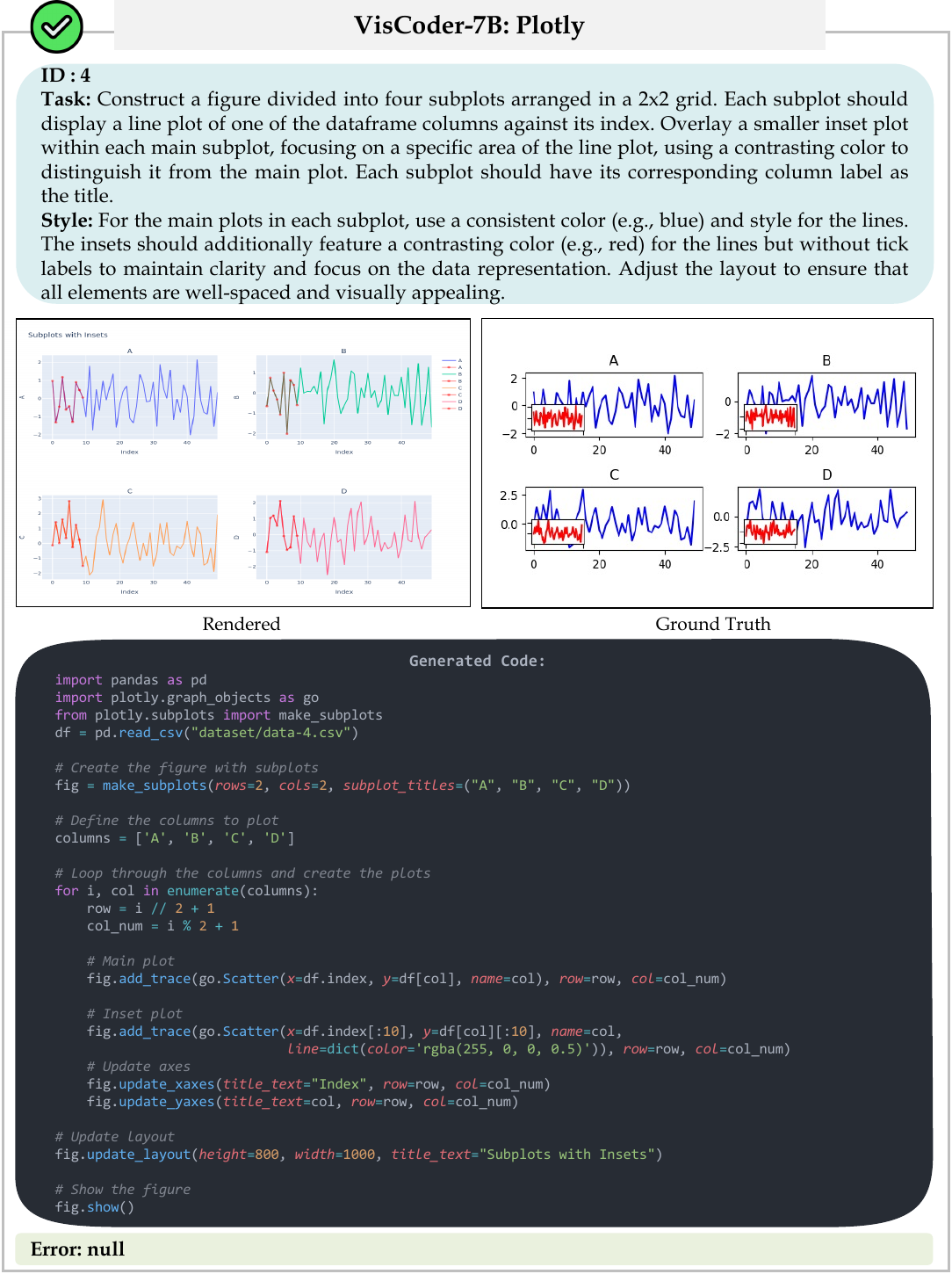}
    \caption{Example of a successful generation in \textbf{Plotly} (ID: 4). The model generates code that executes successfully and produces a plot consistent with the ground truth.
    \newline \centering \hyperref[list:list_of_appendix]{[Back to Appendix Contents]}}
\label{fig:case_plotly_correct}
\end{figure*}

\newpage
\clearpage
\subsection{Plotly: Self-Debug Recovery}
\begin{figure*}[!htbp]
    \centering
    \includegraphics[width=1\linewidth]{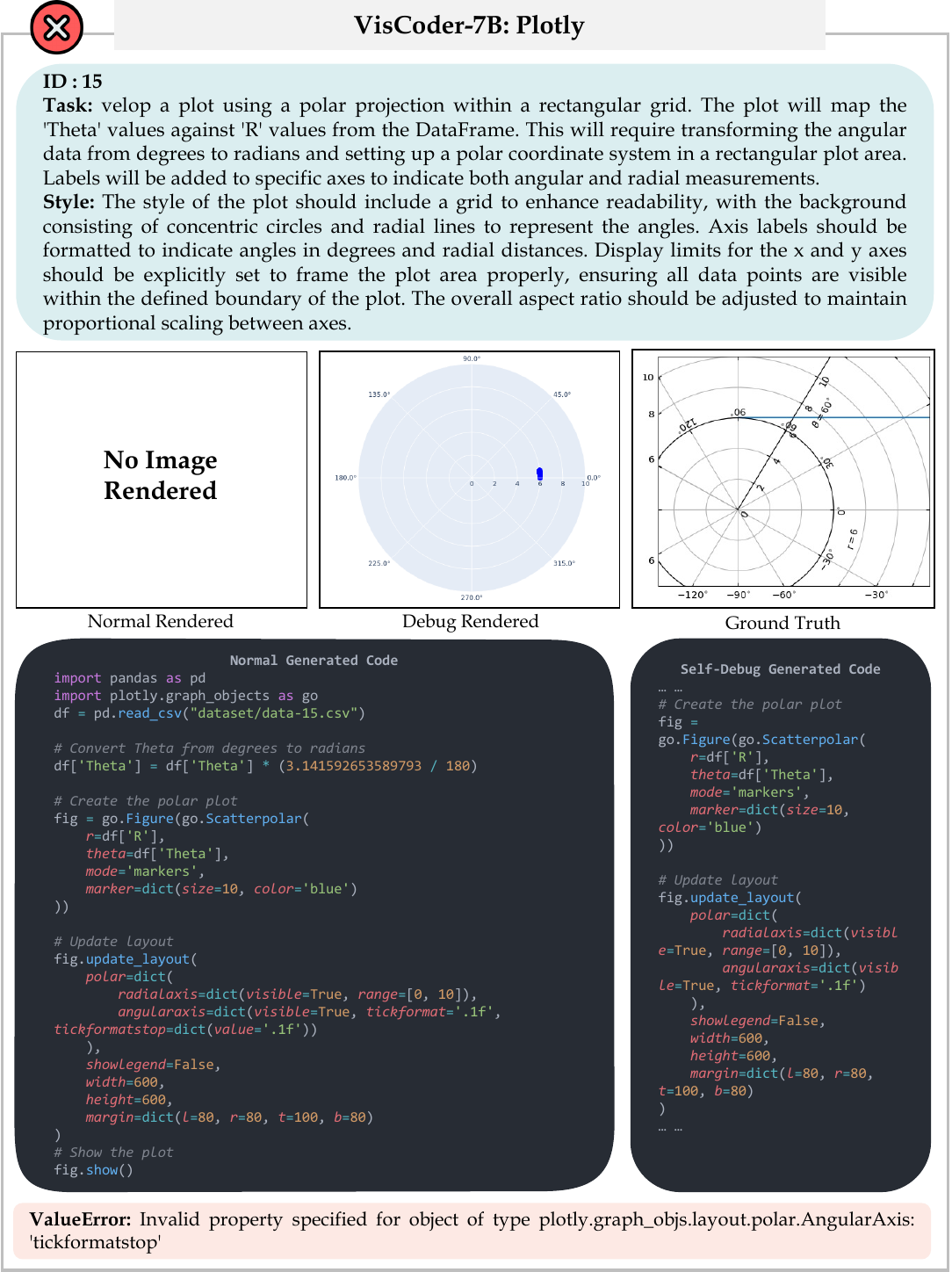}
    \caption{Example of a failed generation in \textbf{Plotly} (ID: 15), where the initial code raises a \textbf{ValueError} and is resolved in the \textbf{Second} round of self-debug, resulting in a corrected plot that matches the intended semantics.
    \newline \centering \hyperref[list:list_of_appendix]{[Back to Appendix Contents]}}
\label{fig:case_plotly_debug}
\end{figure*}